\documentclass[11pt]{article}

\usepackage{amsmath}
\usepackage{amssymb}
\usepackage{slashed}
\usepackage{xcolor} 
\usepackage{graphicx}
\usepackage{makecell}

\usepackage{dcolumn} 
\usepackage{bm} 
\usepackage{epsfig}
\usepackage{epstopdf}
\usepackage{grffile}
\usepackage{color}
\usepackage{colordvi}
\usepackage{rotating}
\usepackage{lscape}
\usepackage{cite}
\usepackage{float}
\usepackage{hyperref}
\usepackage{multirow}

\newcommand{\ie}{{\it i.e.}}

\newcommand{\GeV}{{\rm\ GeV}}

\newcommand{\TeV}{{\rm\ TeV}}

\newcommand{\inab}{\,{\rm ab}^{-1}}
\newcommand{\infb}{\,{\rm fb}^{-1}}

\def\Re{{\cal R \mskip-4mu \lower.1ex \hbox{\it e}\,}}
\def\Im{{\cal I \mskip-5mu \lower.1ex \hbox{\it m}\,}}
\def\ie{{\it i.e.}}

\def\tev{\,{\ifmmode\mathrm {TeV}\else TeV\fi}}
\def\gev{\,{\ifmmode\mathrm {GeV}\else GeV\fi}}
\def\mev{\,{\ifmmode\mathrm {MeV}\else MeV\fi}}

\begin{document}

\begin{center}

\vspace*{15mm}
\vspace{1cm}
{\Large \bf Pinning down the gauge boson couplings in $WW\gamma$ production using forward proton tagging }

\vspace{1cm}

{\bf Seddigheh Tizchang and Seyed Mohsen Etesami }

 \vspace*{0.5cm}

{\small  School of Particles and Accelerators, Institute for Research in Fundamental Sciences (IPM) P.O. Box 19395-5531, Tehran, Iran } \\
\vspace*{.2cm}
\end{center}

\vspace*{10mm}

%
%
\begin{abstract}\label{abstract}

In this paper, we explore the potential of the LHC to
measure the rate of $\mathrm{p}\mathrm{p}\rightarrow  \mathrm{p}~ WW\gamma~\mathrm{p}$ process, also to probe the
new effective couplings contributing to the $WW\gamma$
and $WW\gamma\gamma$ vertices. The analysis is performed at
the $\sqrt{s}=13$ TeV, in the di-leptonic decay channel, and assuming
300 $\infb $ integrated luminosity (IL). In addition to the presence of two
opposite sign leptons, a photon, and missing energy, the distinctive
signature of this process is the presence of two intact protons flying
few millimeters from the initial beam direction in both sides of
interaction points which suppress the background process
effectively. To exploit this feature of signal we benefit from forward detectors (FDs) placed about 200 meters from the interaction point to
register the kinematics of tagged protons. In order to overcome the major sources of backgrounds, we introduced three categories of
selection cuts dealing with objects that strike the central detector,
protons hitting the FDs, and correlations of central
objects and protons, respectively. We also evaluate the probability of
pile-up protons to be tagged in the FDs as a function of the mean
number of pile-up. Then the sensitivity of the LHC to observe this process
and constraints on multi-boson effective couplings are extracted. The
obtained expected limits show very good improvements for dimension-8
quartic couplings and competitive bounds on dimension-6 anomalous
triple couplings w.r.t the current experimental limits. Therefore, we
propose this process to the LHC experiments as a sensitive and
complementary channel to study the multi-gauge boson couplings.
\end{abstract}

\newpage

\section{Introduction}\label{sec:intro}
In the standard model (SM) framework, the non-abelian property of $SU(2)_L\times U(1)_Y$ gauge theory predicts triple and quartic gauge boson self-interactions. 
The triple and quartic couplings are connected to electroweak symmetry
breaking and explore the non-abelian gauge structure. Therefore, studying these couplings provides an important confirmation to the SM. Additionally,
any deviation from the SM prediction of these gauge self-interactions
could be a hint to beyond the SM. For instance, triple gauge boson couplings could be a
consequence of integrating out of non-standard heavy particles at the loop
level while the exchange of heavy new particles at tree level can contribute
to quartic gauge couplings. As a result, any deviation from the SM
predictions observed by the current experimental precision might
appear at quartic rather than triple gauge couplings. From
the theoretical point of view, such deviations can be explained in the
effective field theory framework with high-dimensional model-independent operators that modify the self-interaction of electroweak
gauge bosons or lead to new vertices both known as anomalous
couplings~\cite{Degrande:2012wf}. The anomalous triple gauge couplings (aTGCs) and anomalous quartic gauge couplings (aQGCs) can be
parametrized with dimension-6~\cite{{Belanger:1992qh},{Buchmuller:1985jz}} and dimension-8 operators~\cite{Baak:2013fwa}.

The direct probe of triple and quartic gauge boson interactions could be achieved by measurements of multi-boson productions at the colliders that have been carried out both in the experimental measurements and
phenomenological studies. For instance, the observed bounds on aTGCs
have been obtained from $WW$ production at LEP  with center-of-mass energies $\sqrt{s}=130-209\GeV$\cite{Schael:2013ita:lepcomb}  which  is also measured in leptonic final states at Tevatron~\cite{Abazov:2012ze:d0comb}. Besides, $WZ$ production at semi-leptonic final state at Tevatron with $\sqrt {s}=1.96\TeV$ has been tested for aTGCs~\cite{Aaltonen:2012vu}. Recently, various measurements on the di-boson production such as  $WW$,$WZ$ and $W\gamma$   are performed at the LHC with $\sqrt {s}=7, 8 \TeV$ \cite{{Aaboud:2017cgf:wv8atlas},{Khachatryan:2016vif:wgamma8tevcms},{Chatrchyan:2013yaa:cmsww7tev},{Sirunyan:2017bey:wvcms8}},
   also the same final state in one of the W boson decaying leptonicly
   and the other W or Z boson decaying hadronically has been explored
   at $\sqrt {s}=13 \TeV$
   \cite{{Sirunyan:2019ksz:wzcms13},{Sirunyan:2017ret:ssww13cms}, {Sirunyan:2020tlu:zgacms13},{Sirunyan:2019dyi:w13cms}}. The observed limits on aQGCs in the context of dimension-8 are
  available from exclusive $W$ pair production
 at Tevatron~\cite{Abazov:2013opa}. The constraints on aQGCs at the LHC for same sign W pair
  production plus two jets in leptonic decay of $W$ bosons at
  $\sqrt{s}=8\TeV$ \cite{Aad:2014zda} and $WV\gamma$ production with
  $V=W,Z$ following by semi-leptonic decay of massive gauge bosons at
  the same center-of-mass energy \cite{Aaboud:2017tcq} have been
  obtained. Furthermore, there are many phenomenological studies at lepton colliders such as $e^-e^+$~\cite{{Dawson:1996aw},{Rahaman:2017aab},{Choudhury:1994nt},{Ananthanarayan:2011fr},{Ananthanarayan:2014sea},{Rahaman:2017qql},{Eboli:1993wg},{Choi:1994nv},{Atag:2003wm},{Rizzo:1999xj},{Eboli:1995gv},{Poulose:1998sd},{Stirling:1999ek},{Belanger:1999aw}} and
  hadron colliders~\cite{{Yang:2012vv},{Belyaev:1998ih},{Eboli:2000ad},{Eboli:2003nq},{Baur:2000ae},{Chiesa:2018lcs},{Chiesa:2018chc},{Chapon:2009hh},{Etesami:2016rwu}}
    in which the potential to probe multi-gauge boson couplings have been explored. 
  Beside direct constraints, indirect searches on aTGCs have been
  performed using the data from rare B-meson
  decays~\cite{Bobeth:2015zqa}, 
   as well as coupling measurements of  Higgs boson to electroweak
  gauge  boson at Tevatron and
  LHC~\cite{{Corbett:2013pja},{Dumont:2013wma},{Masso:2014xra}}.
   An alternative possibility to explore the multi-gauge
     boson couplings is via the central exclusive production
     (CEP) at the LHC.
      The CEP happens through the exchange of two color-singlet states
      radiated from two crossing protons resulting in an isolated central system and
      two intact protons that fly in the very forward-backward regions
      of the interaction point. Benefiting from FDs such as
      CMS-TOTEM Precision Proton Spectrometer
      (CT-PPS)\cite{Albrow:2014lrm} and ATLAS Forward Proton (AFP)
      \cite{Grinstein:2016sen} these two protons can be tagged and
      consequently, CEP could be distinguished from inclusive background processes. Therefore,
      exploring CEP processes can provide a unique window to search for new physics namely anomalous gauge couplings.  
The  detailed analysis of central exclusive processes with $W$ boson pair
production including aTGCs and aQGCs have been performed at Tevatron
\cite{Abazov:2013opa}. Moreover,  observed bounds on dimension-6 and
-8 operators via central exclusive
$WW$ production at $\sqrt{s}=7\TeV$, $8\TeV$, and their combinations
without proton tagging are reported by CMS and ATLAS
experiments~\cite{{Chatrchyan:2013akv},{Khachatryan:2016mud:exclwwcms},{Aaboud:2016dkv}}. Additionally,
several phenomenological studies estimated the potential of the LHC for CEP
processes to probe the anomalous gauge boson couplings that can be
found in  Refs~\cite{{Senol:2014vta},{Pierzchala:2008xc},{Chapon:2009hh},{Gupta:2011be},{Monfared:2016vwr}}.
In this paper, we propose $WW\gamma$ production via the CEP as a new
channel at the LHC and explore the potential of this process to probe
 aTGCs and aQGCs. This process is purely
sensitive to gauge boson couplings and can be a complementary channel
to increase the sensitivity to the SM and anomalous $WW\gamma$ and
$WW\gamma\gamma$ couplings. We consider the fully leptonic decay of W bosons  and $300\infb$ of
proton-proton collision data at the center-of-mass energy of $13
\TeV$.
The paper is organized as follows: In section \ref{sec:flux},  a
general description of the flux of emitted photon from a proton is
provided. Section \ref{eff:lag} gives a short review of the effective
field theory approach for anomalous gauge couplings. In section
\ref{sec:wwgama} the CEP of $WW\gamma$ at the LHC is
explained. In Section~\ref{sec:analysis} the strategy of analysis to the optimum
selection of signal and suppression of different sources of
backgrounds are described. Section~\ref{sec:smsens} describes the potential of the LHC to measure the SM $WW\gamma$ process via photon-photon scattering. In
section~\ref{sec:limits} the expected limits on aTGCs and aQGCs are
explained. Finally, in section \ref{sec:conclu} the summary and conclusion are provided.

\section{Photon-photon interaction at the LHC} \label{sec:flux}
 Photon-photon interactions at the LHC can be studied through the
CEP that is defined as
 \begin{equation}
 \mathrm{p} \mathrm{p}\rightarrow \mathrm{p}\oplus X\oplus\mathrm{p}.
\label{eq:cep}
 \end{equation}
 In these type of processes two incoming protons are collided via
 exchange of two color-singlet states such as photons and they remain
 intact. The amount of missing energy of each proton which is carried by each photon, produce state $X$ which can be detected at the central
 detectors while the two unbroken protons fly in the forward and
 backward regions of the central detector with very small angle w.r.t
 their original directions. Therefore, one sees the large rapidity gaps ($\oplus $)
 among the centrally produced state and two forward protons
which is one of the distinctive signatures of  the CEP processes.\par
Despite the fact that the cross sections of the CEP processes are small
w.r.t parton-parton initial state processes, they can be
 measured accurately in a very clean environment due to several reasons. For instance, due to
 the absence of proton remnants, one could obtain the clean
 experimental environment like
 electron-positron colliders. Unlike the usual hard proton-proton
 scattering, by
 measuring the fractional energy loss of each proton ($\xi_{1},\xi_{2}$) which is
 defined as $\xi=\frac{E_\mathrm{p}-E_{\mathrm{p'}}}{E_\mathrm{p}}$ ($E_\mathrm{p}$ and $E_\mathrm{p'}$ are energy of
 incoming and outgoing protons, respectively)  the
 scale of collision can be determined event-by-event basis.  Also
 measurement of the forward protons permits to  predict the
 kinematics of centrally produced state and  matching them  can lead to several orders of magnitude suppression in the background processes.
 Benefiting from  these properties which FDs  granted us,
 the CEP could provide a rich testing ground for electroweak and QCD sector of
 the SM and unique window to physics beyond the SM.\par

\begin{figure}
	\center
	\includegraphics[width=0.6 \linewidth]{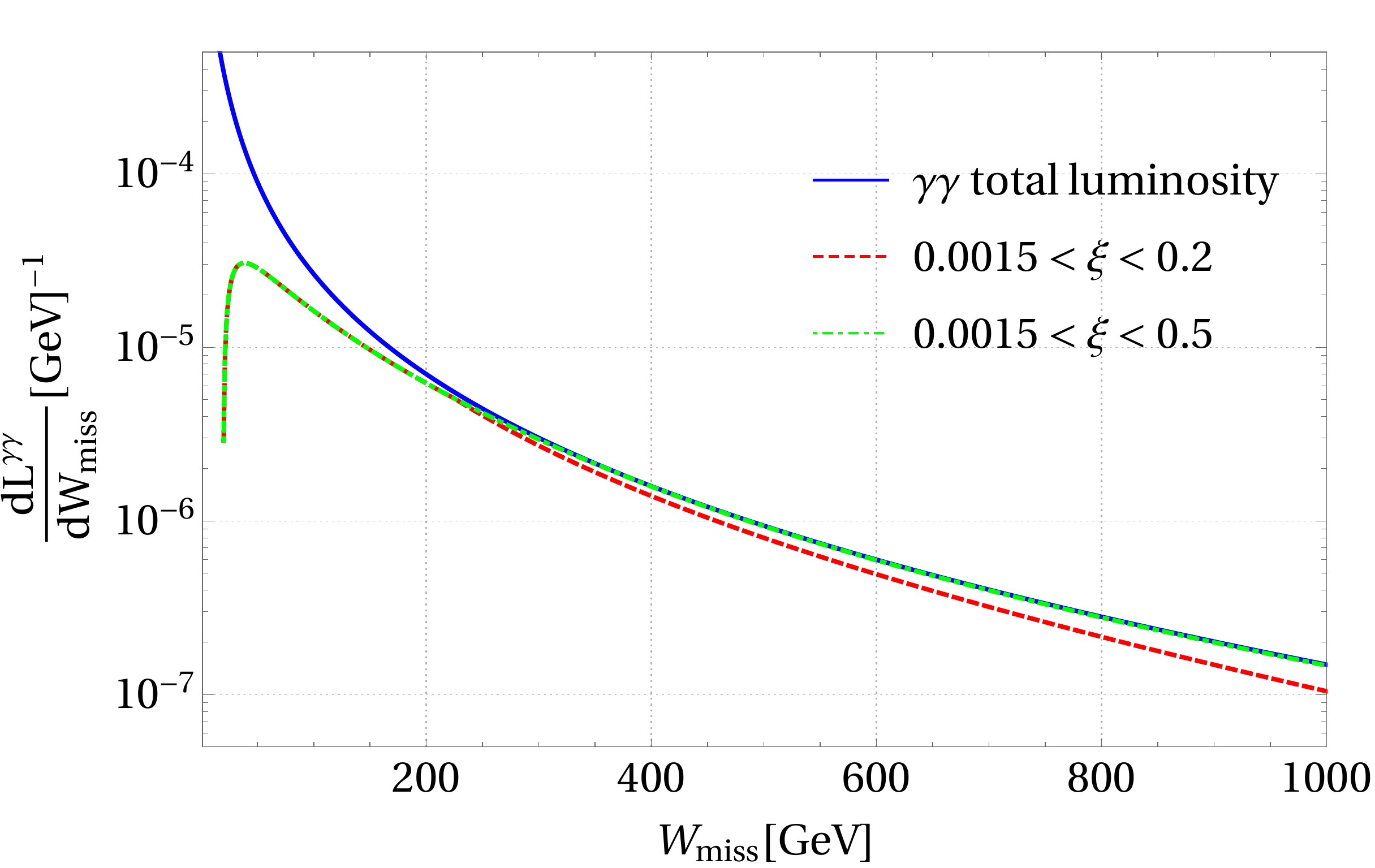}
	\caption{\small Photon-photon luminosity versus two-photon invariant mass at 13 TeV for total and FD acceptance.  }
	\label{lum}
\end{figure}

 The cross section of the CEP process when two photons exchange, can be
 computed in the framework of the Equivalent Photon Approximation
 (EPA)\cite{{vonWeizsacker:1934nji},{Williams:1934ad},{Budnev:1974de}}.
 In this approximation the cross section can be factorized as following
 \begin{equation}
\mathrm{d} \sigma (\mathrm{p}\mathrm{p}\rightarrow \mathrm{p} X \mathrm{p}) =  \hat{\sigma}(\gamma\gamma \rightarrow
X) \mathrm{d} N_{1}^{\gamma} \mathrm{d} N_{2}^{\gamma} ,
    \label{eq:xsecfac}
 \end{equation}

 where $\hat{\sigma}(\gamma\gamma \rightarrow
X)$ is the Born cross section of state X and 
$dN^{\gamma}$ is the number of emitted photons with virtuality $Q^2$
and energy $E_\gamma$. Then the photon spectrum is given by:
 \begin{equation}
 \mathrm{d}^2 N^{\gamma}= \frac{\alpha_\text{em}}{\pi}\frac{\mathrm{d}E_{\gamma}}{E_{\gamma}}\frac{\mathrm{d} Q^2}{Q^2}
 \left[ \left(1-\frac{2 E_{\gamma}}{\sqrt{s}}\right)\left(1-\frac{Q^2_\text{min}}{Q^2}\right)F_E +
 \frac{2 E_{\gamma}^2}{s}F_M\right],
 \label{eq:flux}
 \end{equation}
 where $\alpha_\text{em}$ is fine structure constant and $\sqrt{s}$ is the center-of-mass
 energy of the proton-proton system and $m_\mathrm{p}$ is proton
 mass. The minimum allowed photon virtuality 
 $Q_\text{min}$, $F_\text{E}$, and $F_\text{M}$ are defined respectively as 
  
 \begin{equation}
 Q_\text{min}^2\equiv m_\mathrm{p}^2 E_{\gamma}^2/[(s-\sqrt{s}E_{\gamma})],~F_\text{M}=G^2_\text{M},~ F_E=\frac{(4m_\mathrm{p}^2G^2_E+Q^2G^2_\text{M})}{(4m_\mathrm{p}^2+Q^2)}\,,
 \label{eq:newera:elmagform}
\end{equation}
 and $F_\text{E}$ and $F_\text{M}$  are
 functions of the electric ($G_\text{E}$) and magnetic ($G_\text{M}$) proton form
 factors. In the dipole approximation \cite{Chapon:2009hh}
 \begin{equation}
 G^{2}_\text{E}=G^{2}_\text{M}/\mu_{\mathrm{p}}^{2}=(1+Q^{2}/Q^{2}_{0})^{-4},
 \end{equation} 
the values of $Q^2_0$ and magnetic moment of protons are $0.71 \GeV^2$
and $7.78$, respectively. Therefore, the flux of photon can be obtained by integrating over photon virtuality as follows
  \begin{equation}
  f(E_\gamma)=\int^{Q^2_\text{max}}_{Q^2_\text{min}}\frac{\mathrm{d}^{2} N^\gamma}{\mathrm{d} E_{\gamma} \mathrm{d} Q^{2}} \mathrm{d} Q^{2},
 \label{sm:flux_q2}
  \end{equation}
  
 where the value of $Q^2_\text{max}$ is set to large enough value 2-4 GeV$^2$.
Thus, the total cross section can be obtained as a convolution of the effective photon fluxes and $\gamma \gamma \rightarrow X$ subprocess matrix elements as follows:
\begin{equation}
\frac{\mathrm{d}\sigma^{\gamma\gamma\rightarrow X}}{\mathrm{d}\Omega}=\int\frac{\mathrm{d} \sigma^{\gamma\gamma\rightarrow X}
	(W_\text{miss})}{\mathrm{d}\Omega}\frac{\mathrm{d} \mathcal{L}^{\gamma\gamma}}{\mathrm{d} W_\text{miss}}\mathrm{d} {W_\text{miss}} 
\label{eq:sm:totcross}
\end{equation}
where $\frac{\mathrm{d} \mathcal{L}^{\gamma\gamma}}{\mathrm{d} W_\text{miss}}$ is the two
photons luminosity spectrum which can be obtained by integrating the
product of
two photon rates $f(E_{\gamma 1})f(E_{\gamma 2})$ over the  energy of photons while the two photons invariant mass or equivalently total missing mass of protons $W_\text{miss} = 2 \sqrt{E_{\gamma 1}E_{\gamma 2}}=\sqrt{\xi_{1}\xi_{2}s}$ remains fix. Fig \ref{lum} shows the  effective luminosity of two-photons in pp collision at $\sqrt {s}=13$ TeV as a function of invariant mass of two photons $W_\text{miss}$. The blue (solid) curve is the luminosity spectrum without any cut on the acceptance of FDs. The red (dashed) and green (dot-dashed) correspond to the $0.0015<\xi<0.2$ and $0.0015<\xi<0.5$ ranges of  FDs acceptance  \cite{{Albrow:2008pn},
	{CERN-TOTEM-NOTE}}, respectively. As can be seen,  the photon luminosity decrease by increasing the invariant mass of photons. 
Moreover, applying the lower cut on FDs acceptance leads to  almost one order of magnitude drop in the photon luminosity at low invariant mass $W_\text{miss}$. However, the upper limits on FDs acceptance have little effect on the value of photon luminosity.
The process that we would like to study is $W$-boson pair associated production with a photon in di-leptonic decay of $W$-bosons, generated through di-photon exchange.
It is worth to mention that the process $\mathrm{p}\mathrm{p}\rightarrow W^+W^-\gamma$ has been measured at the LHC \cite{{Djuvsland:2017wky},{Aaboud:2017tcq}}. In the following we briefly review the related effective Lagrangian to aTGCs and aQGCs.

\section{EFT for anomalous gauge couplings involving photon} \label{eff:lag}
In this section, we focus on overall beyond the SM contributions to the triple and quartic gauge boson interactions. These contributions are described through effective field theory (EFT) approach in which the SM is extended by higher-dimensional operators composing by all possible combinations of the SM fields defined as 
	\begin{equation}
	\mathcal{L}_\text{EFT}=\mathcal{L}_\text{SM}+\sum_{i}\frac{c_i^{(6)}}{\Lambda^2}\mathcal{O}_i^{(6)}+\sum_j\frac{c_j^{(8)}}{\Lambda^4}\mathcal{O}_j^{(8)}+...,
	\end{equation}
        	where $\Lambda$ is the  mass scale of any new
                physics. The EFT is valid only for the energy
                $E\ll\Lambda$, $c_i$ are dimensionless Wilson
                coefficients and $\mathcal{O}_i^{(n)}$ represents the
                dimension $n$ operators. The operators respect the Lorentz symmetry and the SM gauge symmetry $SU(3)_C\times SU(2)_L\times U(1)_Y$. Only even-dimension
                operators can contribute if we require lepton and
                baryon number conservation. Therefore, the leading
                effective operators which give contribution to
                vertices containing  multi-bosons are expected from
                dimension-6 operators. Gauge boson interactions within
                the EFT framework can be expressed as two nonlinear
                and linear approaches. In the nonlinear approach, the SM
                gauge symmetry is conserved and is realized by using
                the chiral Lagrangian parametrization given in Refs
                \cite{{Belanger:1999aw},{Eboli:2003nq}}. In this
                approach triple and quartic gauge boson couplings
                appear as dimension-6 operators. In the linear
                approach, the SM gauge symmetry is broken by means of
                Higgs scalar doublet
                \cite{{Belanger:1999aw},{Eboli:2006wa}}. In this
                parametrization, the quartic gauge boson couplings
                without triple gauge boson coupling  appear as
                dimension-8 operators.\\
                The general dimension-6 operators including a neutral vector boson and two charged vector bosons can be described by ten dimensionless couplings \cite{Degrande:2013rea}. However, the number of operators can be decreased to 6 by imposing charge conjugate (C) and parity (P) invariant. The detail of operators is given in Ref. \cite{Degrande:2013rea}.
In this analysis for the ease of comparison to LEP results, we used the anomalous Lagrangian approach for triple gauge couplings which is known as LEP parametrization. 
The aTGCs defining the interaction of photon and W-bosons are expressed as: 
\begin{equation}
\label{eq:atgc_lagrangian}
\mathcal{L}_{WW\gamma}^{(6)} = 
ie(W_{\mu\nu}^\dagger W^\mu A^\nu - W^\dagger_\mu A_\nu W^{\mu\nu}
+ \kappa_\gamma W^\dagger_\mu W_\nu A^{\mu\nu} +
\frac{\lambda_\gamma}{M_W^2}W^\dagger_{\delta\mu}W^\mu _\nu A^{\nu\delta}),
\end{equation} 
where $W_{\mu\nu}=\partial_\mu W_\nu - \partial_\nu W_\mu$  and $A_{\mu\nu}=\partial_\mu A_\nu - \partial_\nu A$ are the field strength of W-boson and photon after symmetry breaking. The dimensionless parameter $\kappa_\gamma$ and $\lambda_\gamma$  are connected to magnetic dipole and electric quadruple moment, respectively. In the SM, the free parameters are 
$  \lambda_\gamma = 0$ and 
$\kappa_\gamma = 1$. In the rest of the paper we constrain the $\Delta\kappa _\gamma$ defined as $1-\kappa _\gamma$ where $1$ is the SM contribution.\par
In general,  to make  equation (\ref{eq:atgc_lagrangian}) gauge invariant under $SU(2)_L$, we have to consider the quartic and higher multiplicity couplings as well.  
As a consequence, the dimension-6  operators contributing in aQGC with two photons and two $W$ bosons are given as \cite{{Eboli:1995gv},{Baillargeon:1994rs}}:
\begin{eqnarray}
\mathcal{L}^{(6)}_{WW\gamma\gamma} &=& \frac{-e^2}{8} \frac{a_0^W}{\Lambda^2} A_{\mu\nu} 
A^{\mu\nu} W^{+\alpha} W^-_\alpha -\frac{-e^2}{16} \frac{a_C^W}{\Lambda^2} 
A_{\mu\alpha} A^{\mu\beta} (W^{+\alpha} W^-_\beta + W^{-\alpha} 
W^+_\beta).\,\,\,\,\,\,\,\,\,\,\,\,\,\,
\label{eq:anom:lagrqgc}
\end{eqnarray}
   Besides, as mentioned above, the lowest order operator of purely aQGCs in the absence of aTGCs are at dimension-8. The dimension-8 quartic coupling operators  can be explained by three classes: longitudinal, transverse and mixing contributions \cite{{Belanger:1999aw},{Eboli:2006wa},{Baak:2013fwa}}. The longitudinal class includes only covariant derivatives of the Higgs field, $D_{\mu} \Phi$. They are given by two independent operators which result in massive vector bosons couplings (see Ref \cite{Baak:2013fwa} for details).
	The mixing class of operators including both field tensor and $D_{\mu} \Phi$ are addressed by seven operators \cite{Baak:2013fwa}. Some of which leading to $WW\gamma \gamma$ vertex are as follows:
	\begin{eqnarray}\label{eq:lag8-transverse-mix}
	\mathcal{ L}^{(8)}_{M,0}&=&\frac{f_{M,0}}{\Lambda^{4}}Tr[W_{\mu\nu}W^{\mu\nu}]\times[(D_{\beta}\Phi)^{\dag}D^{\beta}\Phi], \nonumber\\
	\mathcal{ L}^{(8)}_{M,1}&=&\frac{f_{M,1}}{\Lambda^{4}}Tr[W_{\mu\nu}W^{\nu\beta}]\times[(D_{\beta}\Phi)^{\dag}D^{\mu}\Phi], \nonumber\\ 
	\mathcal{ L}^{(8)}_{M,2}&=&\frac{f_{M,2}}{\Lambda^{4}}[B_{\mu\nu}B^{\mu\nu}]\times[(D_{\beta}\Phi)^{\dag}D^{\beta}\Phi],\nonumber\\ 
	\mathcal{ L}^{(8)}_{M,3}&=&\frac{f_{M,3}}{\Lambda^{4}}[B_{\mu\nu}B^{\nu\beta}]\times[(D_{\beta}\Phi)^{\dag}D^{\mu}\Phi].
	\end{eqnarray}
	On the top of introduced operators, the transverse class operators, including fully field strength tensor, are also possible \cite{Baak:2013fwa}. We should note that some of  transverse class operators (i.e $\mathcal{ L}^{(8)}_{T,0}-\mathcal{ L}^{(8)}_{T,7}$) also contribute at $WW\gamma \gamma$ production cross section; however  their contribution is small comparing to the mixing class operators. \par
	The Lorentz structure of some of the dimension-8 operators are analogous to  dimension-6 aQGC operators. Moreover, most of available constraints on aQGCs are given in dimension-6 parameters. Hence, it is reasonable to explain dimension-8 operators in terms of dimension-6 operators. Considering $WW\gamma\gamma$ vertex, the direct relation between $f_{M,i}$ couplings   with $i=0,1,2,3$, and $a^W_{0}$ and $a^W_C$ couplings are obtained as follows \cite{Belanger:1999aw}:
\begin{eqnarray}\label{8to6trans}
\frac{a_0^W}{\Lambda^2} &=& -\frac{4 M_W^2}{g^2} \frac{f_{M,0}}{\Lambda^{ 4}} - \frac{8 M_W^2}{{g'}^2} \frac{f_{M,2}}{\Lambda^{ 4}},\nonumber\\ 
\frac{a_C^W}{\Lambda^2} &=& \frac{4 M_W^2}{g^2} \frac{f_{M,1}}{\Lambda^{ 4}} + \frac{8 M_W^2}{{g'}^2} \frac{f_{M,3}}{\Lambda^{ 4}},
\end{eqnarray}
where  $g' = e/\cos \theta_W$, $g = e/\sin \theta_W$ and $M_W$ stands for U(1), SU(2) couplings and mass of W boson, respectively. 
\section{WW$\gamma$ production in $\gamma\gamma$ collisions} \label{sec:wwgama}
\begin{figure}
	\center
	\includegraphics[width=0.8 \linewidth]{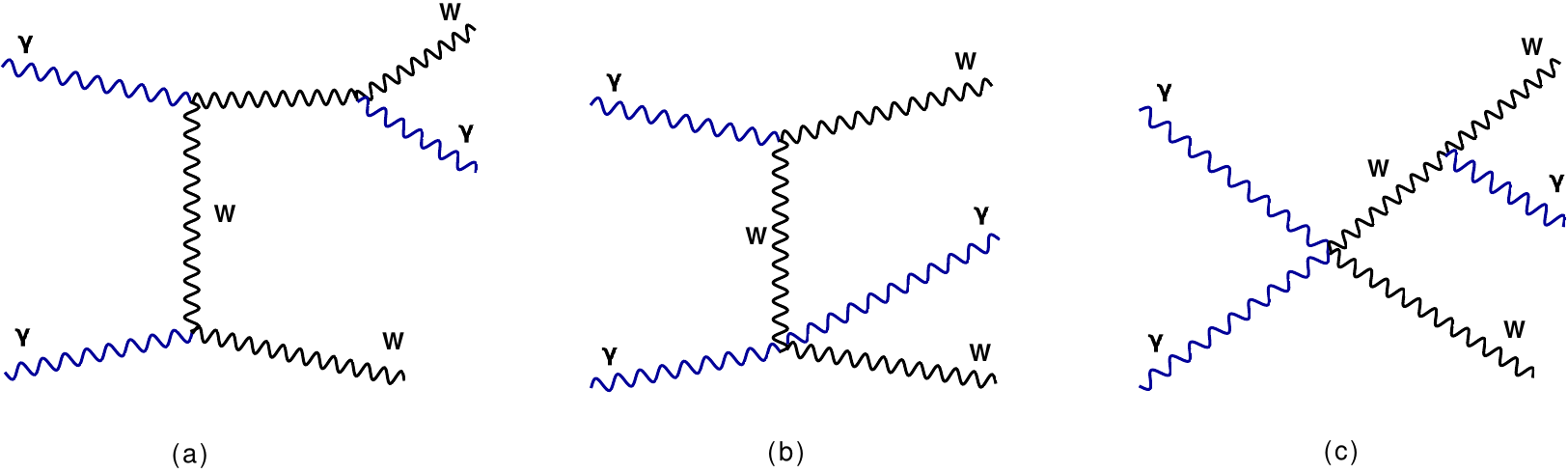}
	\caption{Representative Feynman diagrams for
          $\gamma\gamma\rightarrow W^{+}W^{-}\gamma$
          production at tree level.}
	\label{feynman}
\end{figure}

At the LHC, in addition to the production of $WW\gamma$ via quark-antiquark annihilation, this
process can also  occur via the CEP process as
\begin{equation}
 \mathrm{p}\mathrm{p} \rightarrow \mathrm{p} ~W^+W^-\gamma ~\mathrm{p},
 \end{equation}
where $W^+W^-\gamma$ can be seen in the central detector while the two
intact protons can be detected by the FDs at a long distance from interaction point
and very small angle w.r.t to the proton beams. Figure~\ref{feynman} represents some
of Feynman diagrams for WW$\gamma$ production via the CEP process at tree
level. Diagram (a) has a dominant contribution to the SM
process. Diagrams (b,c) are small at tree level in the SM but
become interesting when one considers aQGCs. In this analysis, we
consider the fully leptonic decay of both W bosons to either electrons or
muons. Therefore, the final state of the  SM signal $WW\gamma$ or $WW\gamma$ process
including the aTQCs and aQGCs  will consist of two opposite sign
leptons, missing energy due to neutrinos from W boson decay and one
isolated photon. We also consider the $\tau$ lepton only if it  decays leptonically 
to electron or muon.
Having the final state of leptons, missing energy, and photon, several sources of background processes will contribute to
our signal region. The background can be divided into two
categories. The first type of backgrounds comes from processes initiated
from the photon-photon interactions such as $l^{+}l^{-}\gamma$,
$\tau^{+}\tau^{-}\gamma$, $WZ\gamma$, $ZZ\gamma$ with $l^{\pm}=e^{\pm}, \mu^{\pm}$. Therefore, in addition to the
similar final state to the signal in the central detector, they have
two intact protons that can be matched kinematically with the central system. The second type of backgrounds which could
contribute are  parton-parton initiated processes,
for instance, $l^{+}l^{-}\gamma$,
$\tau^{+}\tau^{-}\gamma$, $t\bar{t}\gamma$, $WZ\gamma$, $ZZ\gamma$ if
 they coincide with two protons from pile-up.
In this work, we use the \texttt{MadGraph5\_aMC@NLO} package \cite{{Alwall:2011uj},{Alwall:2014hca}} in order
to generate the SM processes and signal processes with aTGCs and aQGCs. In order to
simulate the photon emission from incoming protons for the
processes with photon-photon interaction, the photon PDF based on the
	equivalent photon approximation for low-virtuality
	photons has been implemented in \texttt{MadGraph5\_aMC@NLO}
\cite{Budnev:1974de}. To calculate the total cross section, W-boson mass $m_{W}$= 80.37 GeV and
$G_{F}=1.16639\times10^{-5}  {\mathrm{GeV}^{-2}}$ is considered. 
To calculate the cross section and generate the events of WW$\gamma$ process, including aTGCs and aQGCs, we use the \texttt{FeynRules} package~\cite{feyn} to convert the effective Lagrangian to the UFO model~\cite{ufo} which
can be linked to the \texttt{MadGraph5\_aMC@NLO}. 
\texttt{MadGraph5\_aMC@NLO}  is also used to generate several
photon-photon or parton-parton initiated background processes. To perform parton showering and hadronization, all the generated
samples are passed through \texttt{Pythia 8} \cite{pythia8}. In this
analysis, we also consider the fast simulation of LHC-like detectors to consider the effects of detector response on the final
reconstructed objects using the \texttt{Delphes 3.4.1} package~\cite{deFavereau:2013fsa}.
\section{ Analysis Design} \label{sec:analysis}
In this section, we describe the strategy for analysis including the
object selection cuts, event selections and discuss the contribution of each source of background.
Depending on the flavor of leptons  decayed from W boson, we divide our signal region into the same flavor leptonic (SF) channel consists of $e^{+}e^{-}, \mu^{+}\mu^{-}$ events and different flavor (DF) channel $e^{+}\mu^{-},
e^{-}\mu^{+}$, because the  SF channel suffers from 
a large  contribution of  $l^{+}l^{-}\gamma$ background process while DF
does not. 

\subsection{ Selection cuts} \label{selecut}
\begin{figure}
	\begin{center}
          \resizebox{0.46\textwidth}{!}{\includegraphics{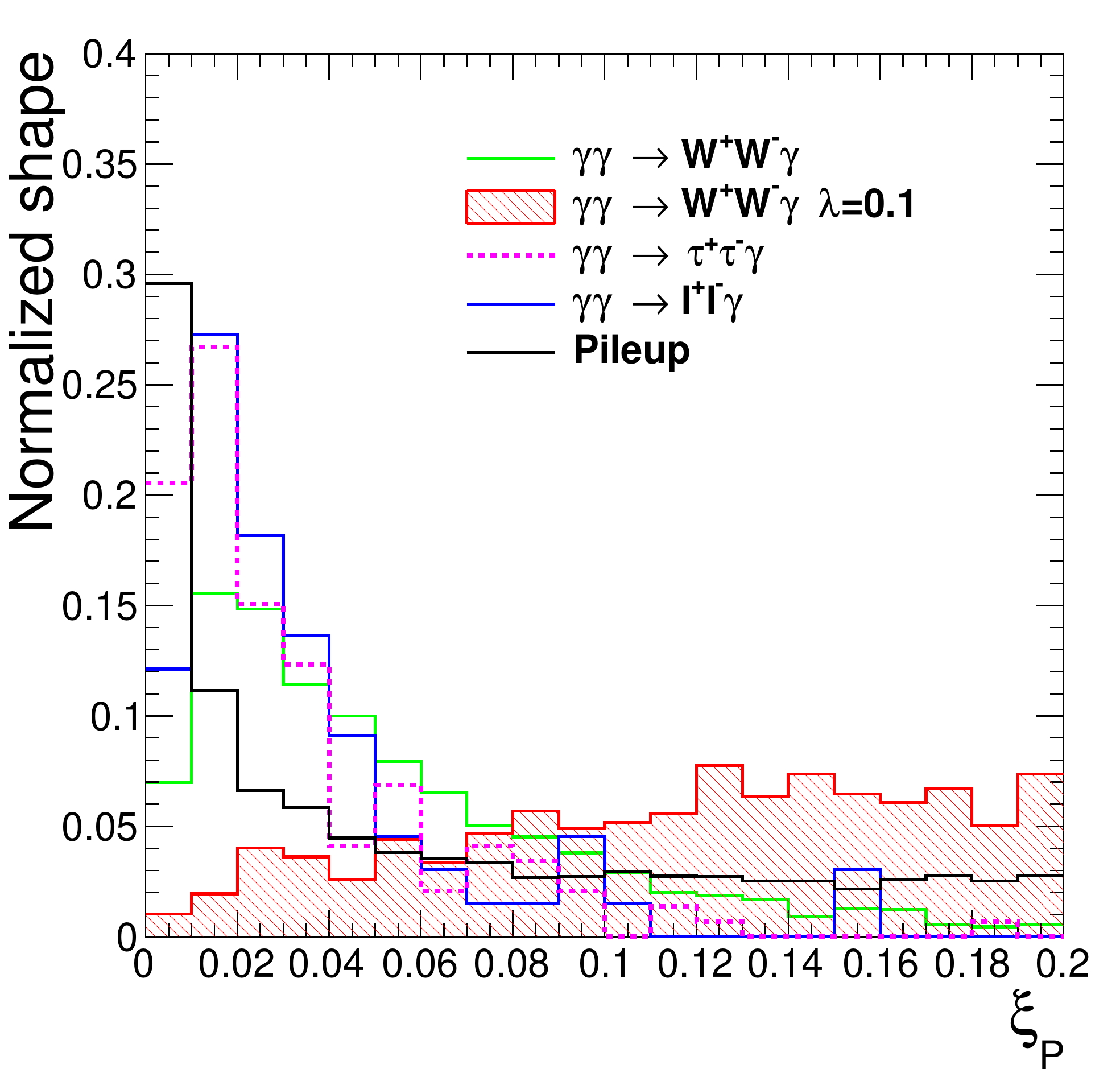}}
            \resizebox{0.46\textwidth}{!}{\includegraphics{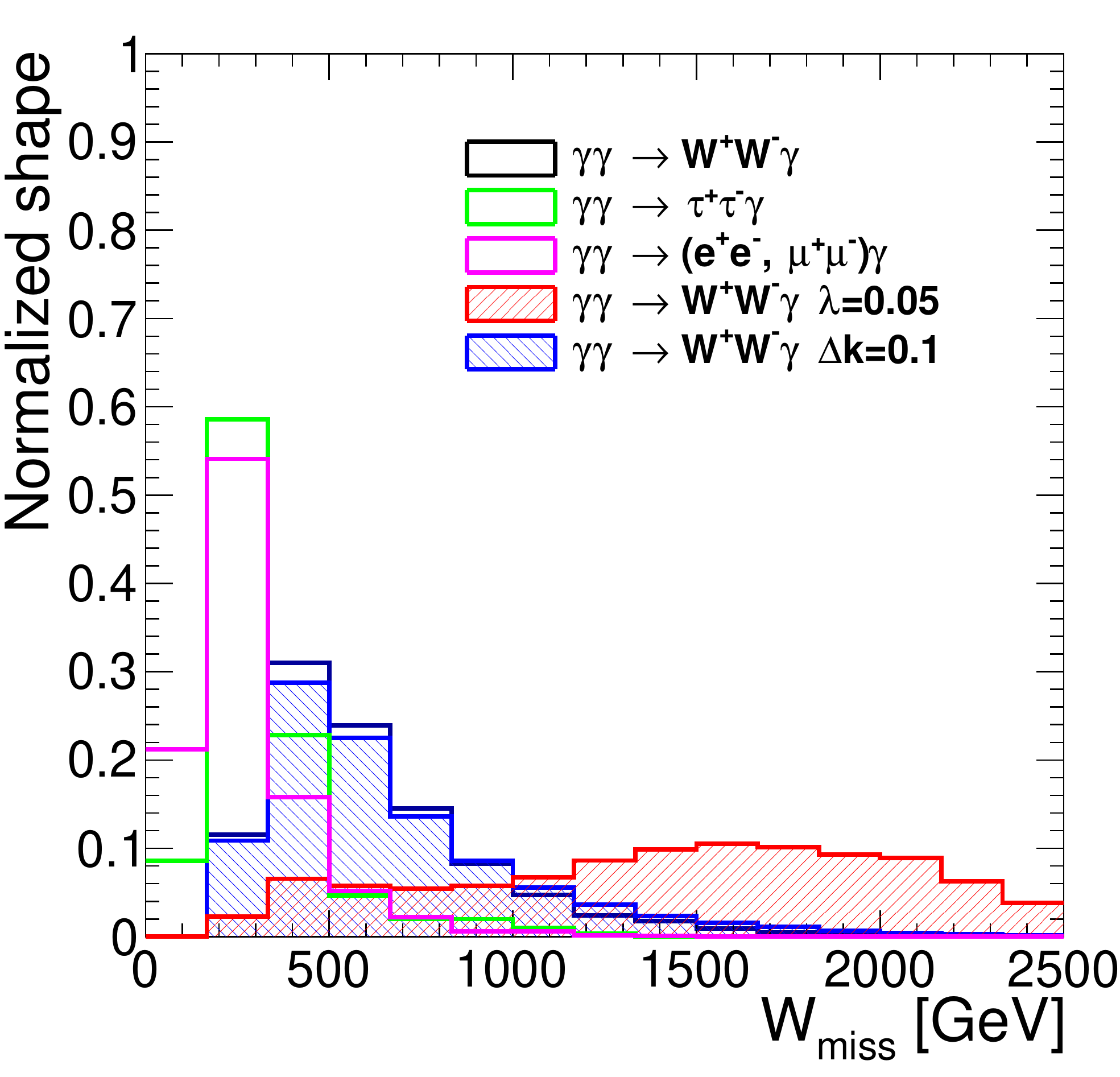}}          
		\caption{ Distribution of fractional energy loss of protons reaching the FDs (left) and 
protons missing mass (right) for SM $\mathrm{p}\mathrm{p}\rightarrow \mathrm{p}W^+W^-\gamma
\mathrm{p}$, $\mathrm{p}\mathrm{p}\rightarrow \mathrm{p}l^+l^-\gamma \mathrm{p}$ and  two $\mathrm{p}\mathrm{p}\rightarrow
\mathrm{p}W^+W^-\gamma \mathrm{p}$ samples including aTGCs correspond to
$\lambda_{\gamma}= 0.05$ and $\Delta \kappa= 0.05$.}\label{kisilostmass}
	\end{center}
\end{figure}

In order to select the $ WW\gamma $ signal events including the SM
or anomalous coupling contributions, we apply three categories of
selection cuts. The first set of cuts includes the central detector requirements which are applied in order to select the objects needed to construct the signal final state events
and suppress the backgrounds, optimally. The second set of cuts is applied to the tagged protons in order to adopt the acceptance of
the FDs. The third type of cuts is beneficial form
the kinematic correlation of central produced state and detected protons in
the FDs to reject non-exclusive backgrounds as well as exclusive
backgrounds with the different expected proton's kinematics according to the signal. 
For the first category of cuts (type I),  we require to have two
opposite sign well-isolated leptons, both of them required to have
$p_{T,l}>10$ GeV but at least one of them must pass $p_{T,l}>20$ GeV
and both are required to be in $|\eta_{l}|<$2.5. It
  should be pointed out that required
$p_{T,l}$ thresholds are totally in agreement with the current
thresholds of double
lepton triggers which are used in the current experiments such as
CMS. Therefore one expects the high trigger efficiency considering
these cut values. In addition, we veto events containing any extra
loose lepton with $p_{T,l}>10$ GeV and $|\eta_{l}|<$ 2.5 in order to suppress the contribution from
$WZ\gamma$, $ZZ\gamma$ background processes. We
also demand to have exactly one isolated photon with $p_{T,\gamma}>
20$  GeV and $|\eta_{\gamma}|<$ 2.5. To suppress the backgrounds
without W boson, we require the missing transverse energy, 	$\slashed{E}>30{\GeV}$. In order to suppress the contribution of events with a photon
radiating from  W and Z decay product as well as preventing the drop of 
reconstruction efficiency due to the close-by lepton from photon, we require
to have $\Delta R_{\gamma, l}=\sqrt{\Delta \phi^2+\Delta \eta^2}>$
0.5. The last criteria to suppress the backgrounds containing jets such
as inclusive $t\bar{t}\gamma$, is to veto events containing more than one
jets with $p_{T,j} > $40 GeV and $|\eta_{j}| < $5.0. The reason to apply 
veto cut on the  number of jets $N_j>2$ is that the probability to reconstruct a jet from pile-up is not negligible in high pile-up conditions, even for purely leptonic signal events. Therefore, to maintain the optimum amount of signal versus
high rejection of backgrounds, we loosen the number of jets in the
veto condition.
In the second type of cuts (type II) we require to have at least one proton in each side of interaction point to
be within the FD acceptance. The acceptance of the detector is
usually expressed in terms of the fractional energy loss of each proton. In this analysis, we consider two scenarios for FD acceptance corresponding to
$0.0015<\xi<0.2$ and $0.0015<\xi<0.5$. However, in order to suppress the different sources of backgrounds, we require $\xi$ to be greater
than 0.008. Figure \ref{kisilostmass} (left) indicates the distribution of
fractional energy loss of both protons reaching the forward
detectors. One could see from Figure \ref{kisilostmass} (left) that applying  a lower cut on the $\xi$ will reduce
the contribution of photon initiated backgrounds as well as pile-up
backgrounds, effectively. Furthermore, we exploit from the high correlation
of primary vertex (PV) displacement in the $z$-direction and arrival time of
both tagged protons to the timing FDs in the CEP processes, to put down
the inclusive background contribution, effectively.
For the third type of requirements (type III), we restrict the protons missing
mass $W_\text{miss}$ to be larger
than 200 GeV. As indicated in Figure~\ref{kisilostmass}
(right) protons missing mass for the CEP $W^+W^-\gamma$ starts from 200
GeV which is approximate energy for the production of two on-shell W
bosons. While for the less heavy state such as the CEP $l^+l^-\gamma$, the
threshold is smaller. Therefore, having this cut effectively reduces
the contribution of photon-photon initiated $l^+l^-\gamma$
background. In addition to
that, one can use the conservation of momentum in the $z$-direction in
order to obtain the missing longitudinal momentum of the central
system. Thus, the central mass can be reconstructed partially and
could be used to reject both the CEP and inclusive backgrounds  when it is
compared to the protons missing mass. This cut will be explained and
discussed in detail in Section~\ref{inclpileup}.
 \subsection{Pile-up Implementation}
Pile-up is referred to the multiple soft proton-proton
interactions in each bunch crossing of the LHC which is happening along
with the hard process coming from the PV. The average number
of pile-up interactions per bunch crossing $<N_{PU}>$ is depended on the condition of
the machine that collides the protons to each other, such as the energy of
protons, the number of protons in each bunch and etc. The mean value of 
pile-up at the LHC varies between 20-50 from Run I to II and the
expectation for high luminosity LHC is between the 140-200. The overall effect of pile-up interaction is the production of soft hadrons that propagate to all layers of detectors and biases the measured quantities and degrade the resolution of the reconstructed final state particles of hard processes.
In order to estimate the effects of the pile-up interactions, we
simulate the minimum bias events using the \texttt{Pythia 8}. Then
the superposition of generated minimum bias events and the PV
of hard processes is performed using the \texttt{Delphes
  3.4.1}. In order to implement pile-up interactions in each
event we take  similar parameters that considered for modeling 
pile-up in the CMS detector at the LHC. In this analysis the
simulation of all the
signal and background samples are performed, considering the average
number of pile-up as $<N_{PU}>$= 30.
In this analysis, the
variables which are affected by the pile-up are isolation of leptons, photons as well as missing energy. Therefore, to alleviate the
effects of pile-up, we subtract the contribution of soft charged
particles originated from the vertices  which are  far enough from the
primary vertex. Furthermore, to remove the contribution of neutral pile-up
the FastJet area method is used
\cite{Cacciari:2008gn}. This method considers the homogeneous energy
density imposed by neutral pile-up particles and the area where the
isolation of leptons and photons are effected to subtract the neutral
contribution of the pile-up.
In addition to the above effects, the presence of the pile-up in the exclusive searches is very important
 as they can produce protons that may lay in
the acceptance of FDs. We will discuss the effects of these detected
protons in the next section. 
 \subsection{Inclusive  backgrounds with Pile-up protons}\label{inclpileup}
One of the main sources of background processes that can contribute to our
signal regions is non-exclusive backgrounds. The main processes which can produce a similar
signature as our signal in the central detector are $t\bar{t}\gamma,
l^+l^-\gamma, W^+W^-\gamma,
W^+Z\gamma, ZZ\gamma$.  The cross sections of these processes are  several orders of magnitudes larger than the signal. They may pass the type I selection requirements. However, they can contribute to our signal region only if they pass also the
type II and III selection cuts. This can
happen if the event of such inclusive processes coincides with
protons in the acceptance of FDs produced by pile-up
interactions.
The main proton-proton processes that may
produce protons in the FDs are elastic and single diffractive processes. Other interactions such as double
diffractive and non-diffractive processes have fewer contributions as
these interaction does not produce intact protons directly and the trapped
protons from these processes can result from the dissociation of incoming protons.
In addition, the elastic interactions suppress heavily due to the lower cut on the proton acceptance region $\xi >$ 0.008, introduced in the previous section,  as the outgoing
protons  are expected to have very small $\xi$. Therefore, the main process
that can produce two detectable protons on each side of FDs is from multiple single
diffractive processes that occur at each proton bunch crossing.
The appearance of protons in the FDs due to pile-up is independent
of inclusive processes and vary only by changing the average number
of pile-up. Therefore, the fraction of events that have at least one
proton in the acceptance region of each side of FDs, as well as events that satisfy the timing requirement, can be calculated as an independent
factor from the specific inclusive processes. Then to obtain the
backgrounds yield after applying the Type II selection cuts, one can
multiply the number of inclusive events that remain after central
selection cuts by these calculated efficiencies. We will calculate and
report the type II selection cut efficiencies for different pile-up
scenarios that can be used for any study including this type of background. 
\begin{figure}
	\begin{center}
		\resizebox{0.48\textwidth}{!}{\includegraphics{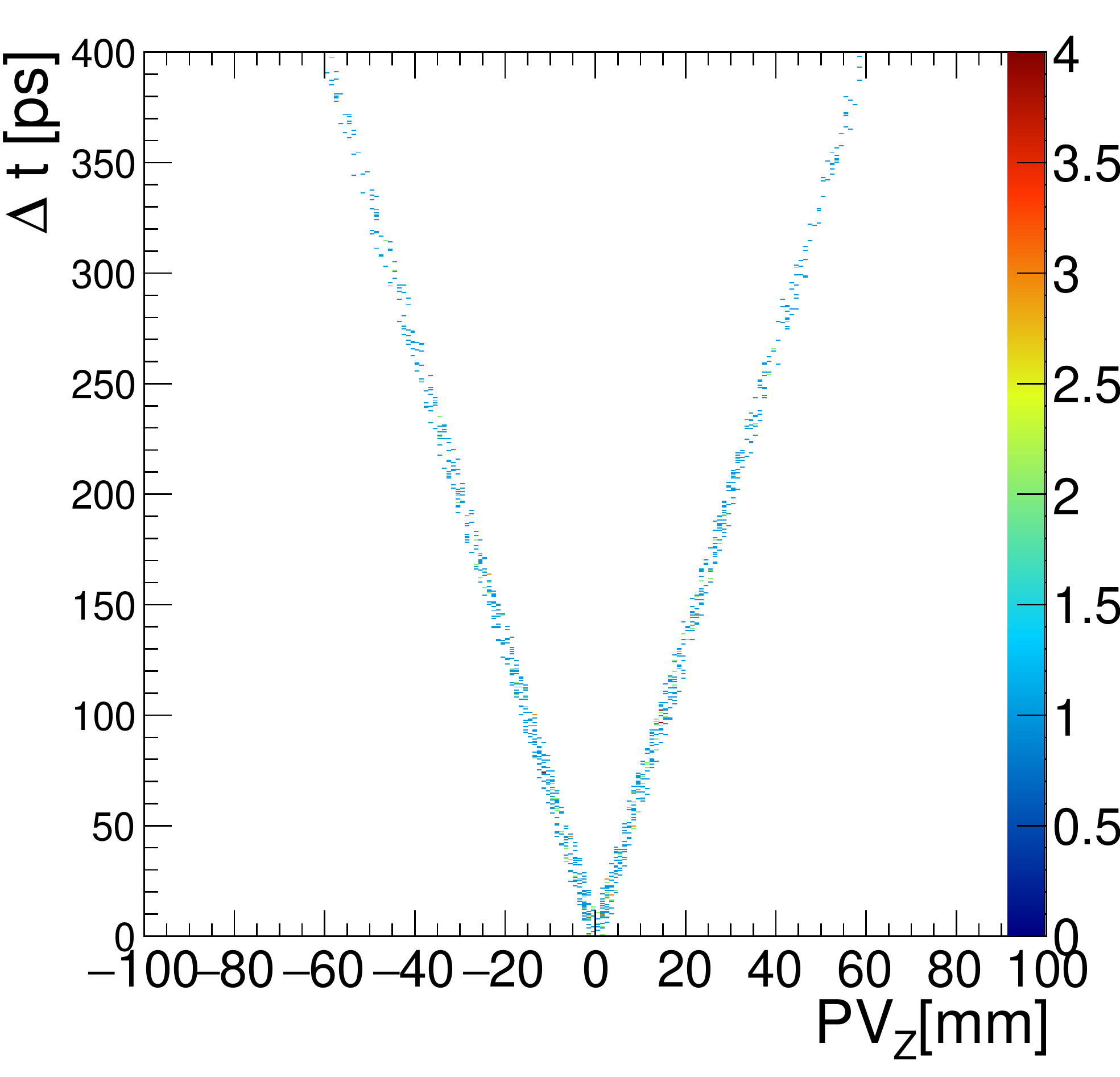}}  
		\resizebox{0.48\textwidth}{!}{\includegraphics{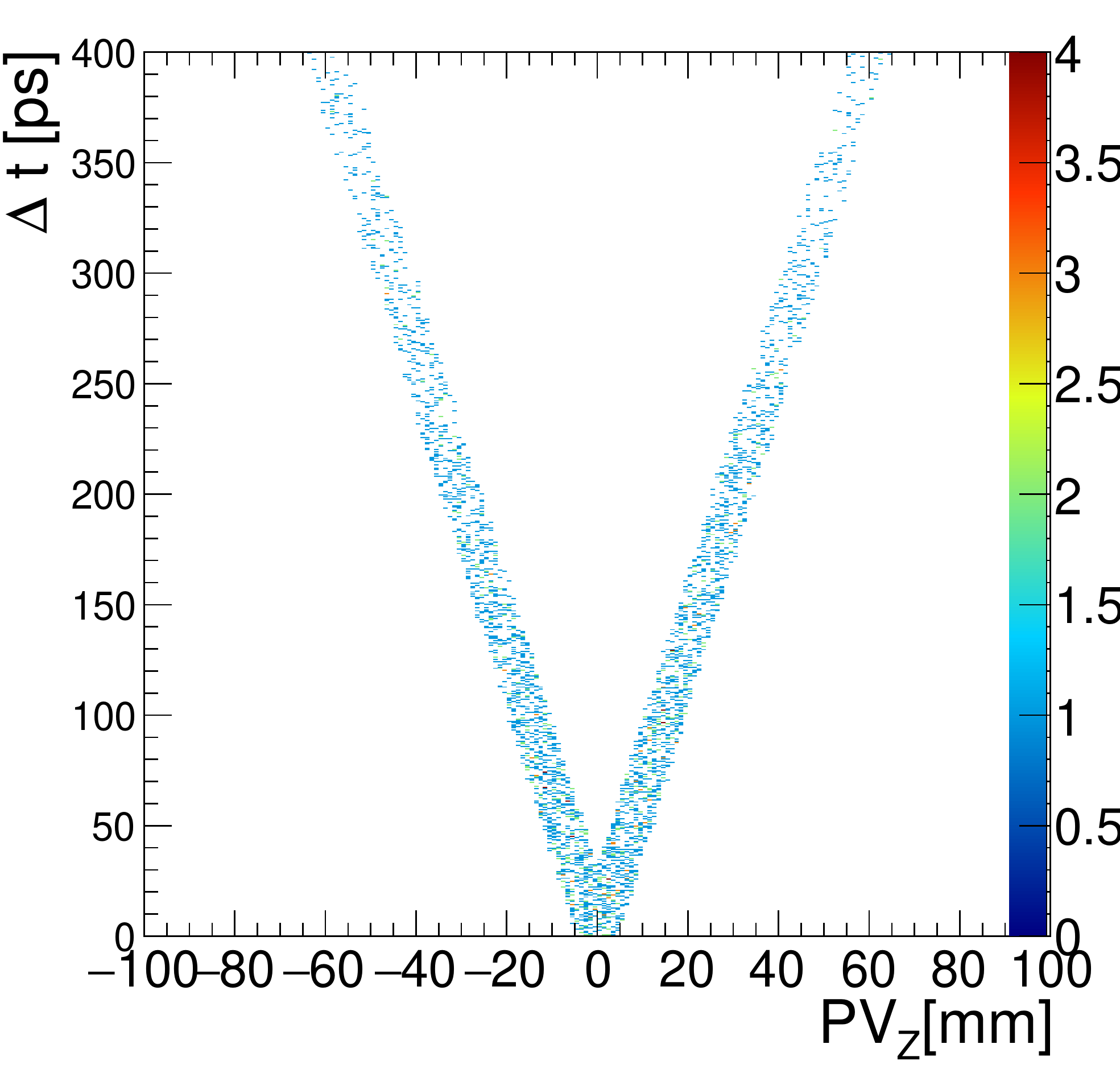}}  
		\caption{Absolute time of flight
                    difference of two protons detected in the timing FDs versus PV position in $z$-direction, 
			  considering 10 ps(left) and 30 ps(right) resolutions. This
			correlation is calculated for the SM 
			process $\mathrm{p}\mathrm{p}\rightarrow
			\mathrm{p}W^{+}W^{-}\gamma \mathrm{p}$.}\label{timewindow}
	\end{center}
\end{figure}
The type II selection cuts described in Section \ref{selecut}
except for the timing requirement which will be
  discussed here.

\begin{figure}
	\begin{center}
		\resizebox{0.48\textwidth}{!}{\includegraphics{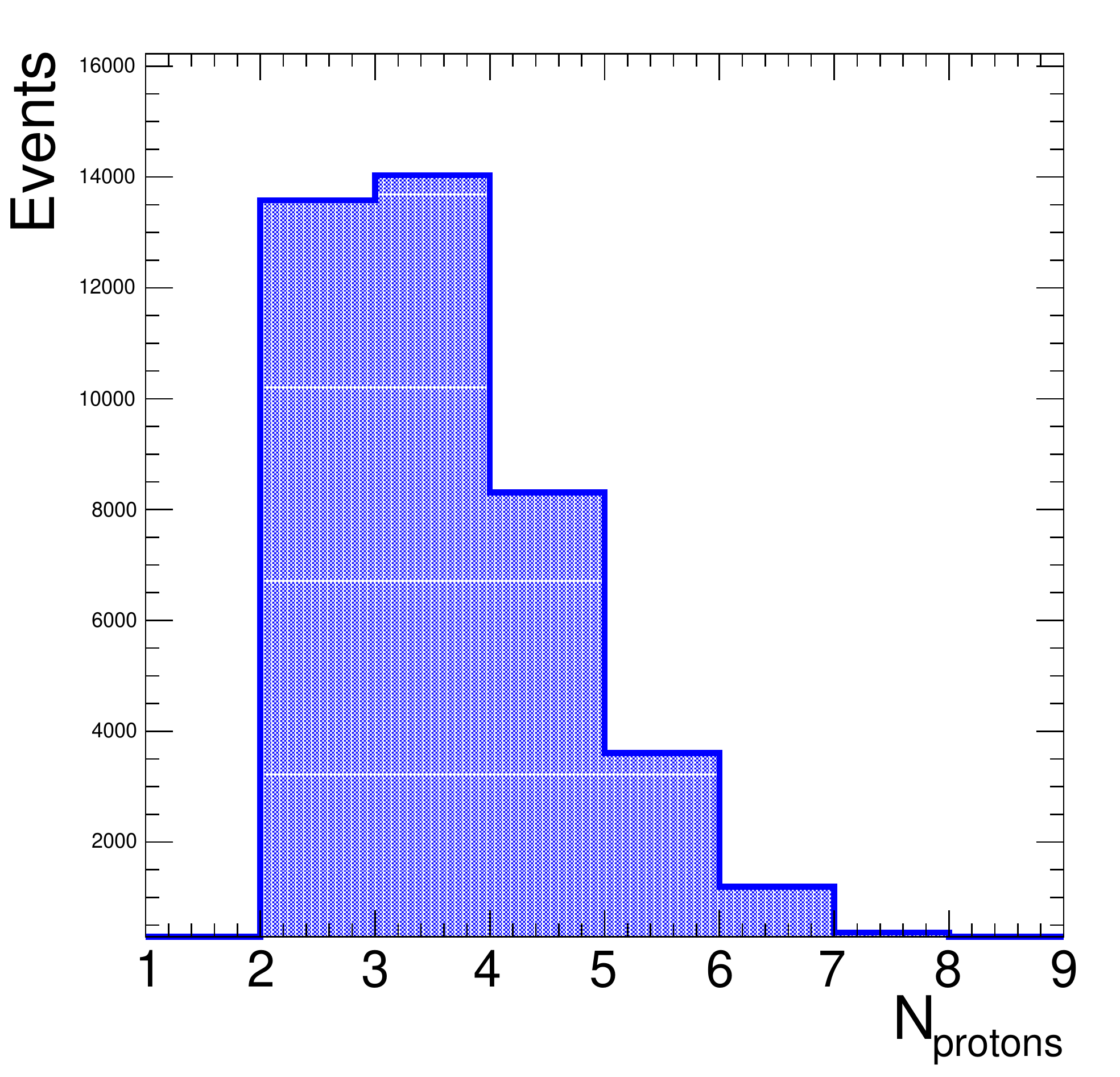}}  
		\caption{ Multiplicity of pile-up protons passing the
                  $0.008<\xi<0.2$ FDs acceptance. The average number of
                  pile-up is $<N_{PU}>$=30 per event. The number of
                generated pile-up events are 100k.}\label{nproton}
	\end{center}
\end{figure}

\begin{figure}
	\begin{center}
		\resizebox{0.48\textwidth}{!}{\includegraphics{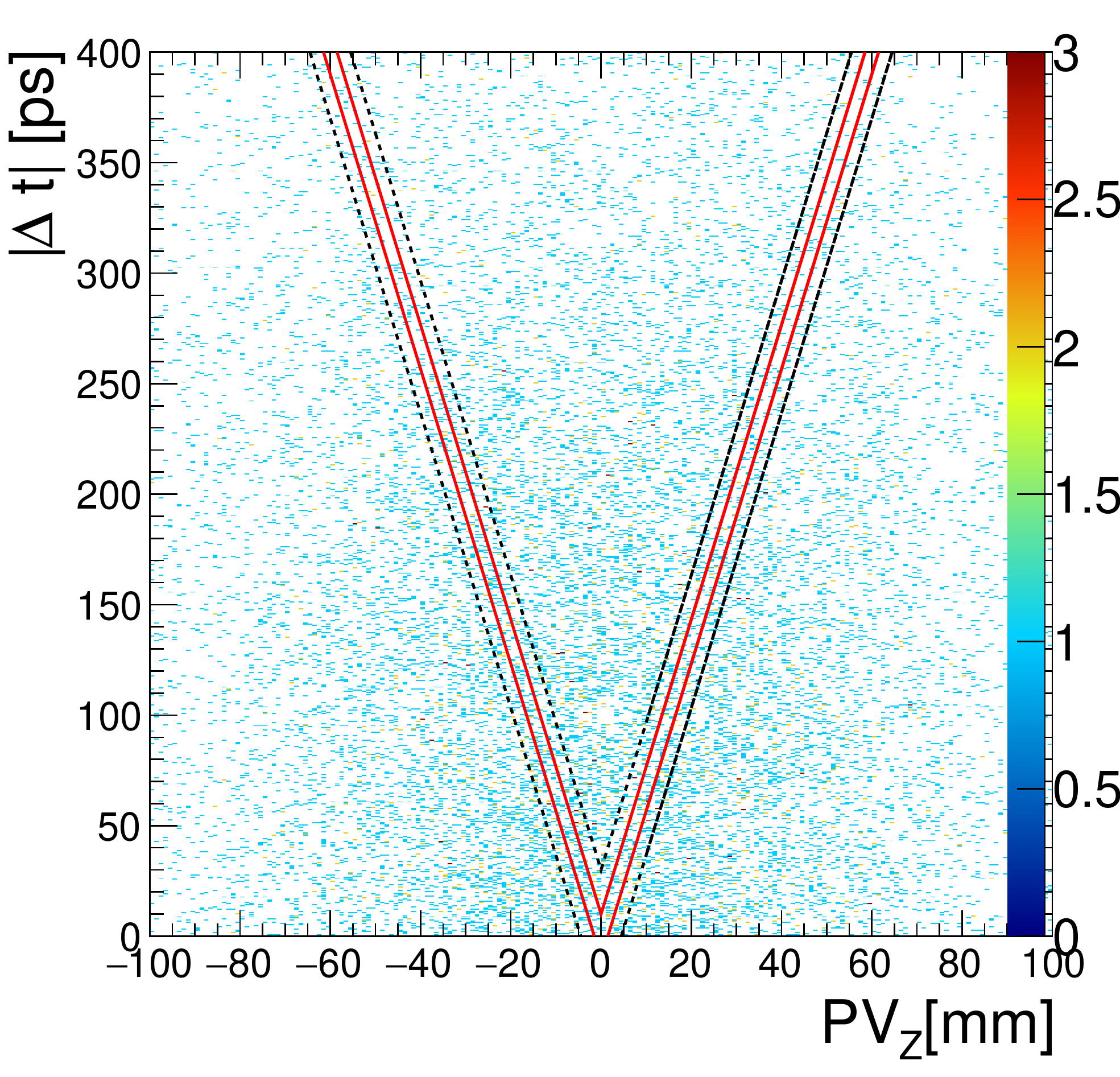}}  
		\caption{Distribution of absolute time difference of superimposed 
				pile-up protons tagged in the FDs
                                versus displacement of PV in the
                                $z$-direction for inclusive
                                $W^+W^-\gamma$ process. The protons
                                are restricted to pass
                   $0.008<\xi<0.2$ requirement. The average
                number of pile-up per each bunch crossing is $<N_{PU}>$=30. The
                events between red
                (black dashed) lines will pass the timing requirement
                due to coincidence of pile-up protons with their PV
                position considering 10(30) ps resolutions.}\label{pileuptime50}
	\end{center}
\end{figure}

In the CEP
processes primary vertex position in the longitudinal direction is
proportional to the difference between the arrival time of two protons
 to the FDs as $z_\text{PV}\propto \frac{(t_{1}-t_{2})}{2}$, while for
inclusive processes superposed by the pile-up protons are
not. Therefore,  depending on the timing resolution of time of
	flight detectors, it could be used to reject the inclusive backgrounds
several orders of magnitudes. The benchmark resolution considered for timing FDs is
between the 10-30 ps \cite{Adamczyk:2015cjy,
  Albrow:2014lrm} corresponds to the uncertainties of
$\sigma_{\text{PV}_{z}}=$ 2.1 and 6.3 mm on the PV position,
respectively. Figure~\ref{timewindow} shows the correlation
between the displacement of the primary vertex in $z$-direction and the absolute difference between the arrival
time of two tagged protons in the FDs for the SM $W^{+}W^{-}\gamma$, CEP
process assuming the 10 ps (left) and 30 ps (right) resolutions for
timing detectors.
inclusive background processes will pass the
  timing cut if the distance between vertex position obtained from
  pile-up protons and the PV position is closer than 2.1 (6.3) mm for
  considered resolutions of 10 (30) ps, respectively. These distance
  values can be obtained simply from timing resolutions of FDs. In order to apply this
requirement, one needs to select one pair of protons out of different possible
combinations. Because the number of pile-up protons
that reach the timing FDs can be exceeded from two. Figure~\ref{nproton} indicates the multiplicity of
protons passing the acceptance cuts. 
Therefore, we require to select a
pair of protons  which is placed in the closest distance w.r.t to the
PV  by defining the $\delta r=
\sqrt{(Z_\text{PV}-Z_{\mathrm{p}1})^{2}+(Z_\text{PV}-Z_{\mathrm{p}2})^{2}}$ where, the $Z_{\mathrm{p}1,\mathrm{p}2}$
are vertex position of each tagged protons, and $Z_\text{PV}$ is the vertex
position of PV in the $z$-direction. Thus, a pair of protons with the smallest value of
$\delta r$ in each event is selected. Figure~\ref{pileuptime50} shows the distribution
of PV position of $W^+W^-\gamma$ inclusive process versus absolute difference arrival time of two selected pile-up
protons.
The area enclosed between the red  (black dashed) lines represents accepted inclusive events that the time difference of the arrival of their tagged protons falls within the uncertainty range of the PV position imposed by the timing detector resolutions of 10(30) ps.
Table~\ref{doubletag} shows the calculated fraction of events remained after applying the acceptance cut as well as the time of flight (TOF) cuts for the average number of pile-up from $<N_{PU}>$=10-140, assuming two scenarios for protons
acceptance. It is clear from Table~\ref{doubletag}
  that the probability of pile-up protons tagged in the FDs increases with rising the average number of pile-up $<N_{PU}>$. Therefore, it is necessary to pay careful
  attention to the inclusive background processes along with the pile-up
  protons in the high pile-up condition to estimate the realistic
  background contribution.

The last category of selection cuts  which we call it type III, depends on both the
central state and tagged proton kinematics. The first requirement is a lower
cut on the $W_\text{miss}$ which is explained in the
section~\ref{selecut}. In addition to that, in the CEP processes, one
can reconstruct the mass of central system from the fractional
energy loss of both tagged protons $W_\text{miss}$, whereas it is not true if the
protons come from pile-up interaction, explained earlier in this
section. 
\begin{table}
	\begin{center}
		\begin{tabular}{|l|l|l|l|l|l|l|}
			\hline
			\multicolumn{7}{|c|}{Double tagging efficiency} \\
			\hline
			\hline
			&$< N_{PU}>$ & 10& 30& 50 & 100 &140  \\ \hline                                      
			\multirow{6}{*}{Type II } &	$0.0015<\xi<0.2$  & 0.06 &0.31 &0.56 &0.87 &0.95 \\ 
			& $0.0015<\xi<0.5$&0.30 &0.81 & 0.95& 0.99&1.00 \\ \cline{2-7}
			& \multirow{2}{*}{$\xi >0.008$ }    &0.03   & 0.19& 0.38&0.73  &0.86\\
			&&0.26   &0.76 & 0.93&0.99  &1.00\\ \cline{2-7}
			& \multirow{2}{*}{TOF }    &0.001   & 0.008& 0.016&0.036  &0.048\\
			&&0.010   &0.04 & 0.06&0.011  &0.157\\ \hline
		\end{tabular}
	\end{center}
	\caption{The fraction of remained events assuming the FDs	acceptance, lower cut on the $\xi_{1,2}$, and TOF
		requirement for different average
		numbers of pile-up are reported. At least one proton on each
		side of interaction point is required. The obtained
		fraction of events at each level includes all other previous
		cuts. The efficiency of double proton tagging is increased by rising the number of pile-up.} \label{doubletag}
\end{table}
Therefore, in the case of full reconstruction of the central
system, there will be a direct correlation between $M_{X}$ with $X=l_1 l_2 \gamma (\nu_1+\nu_2)_\text{rec}$ and $W_\text{miss}$ which results to significant rejection of backgrounds. On the other hand, there is a source of ambiguity in the di-leptonic channel of the central system of our interest CEP $W^{+}W^{-}\gamma$ (either the SM or including the anomalous
couplings) process, due to the presence of two
neutrinos. Regarding the neutrinos, the only known information from
central detector is the sum of
missing energy in $x$ and $y$ directions, while the presence of FDs allows us to
obtain the total missing momentum in the $z$-direction via
conservation of momentum in the longitudinal direction 
\begin{equation}
  p_{z}(l_{1})+p_{z}(l_{2})+p_{z}(\gamma)+p_{z}(\nu_{1}+\nu_{2})=p_{z}(\mathrm{p}_{1})+p_{z}(\mathrm{p}_{2}).
 \label{eq:zbalance}
\end{equation}
\begin{figure}
	\begin{center}
          \resizebox{0.45\textwidth}{!}{\includegraphics{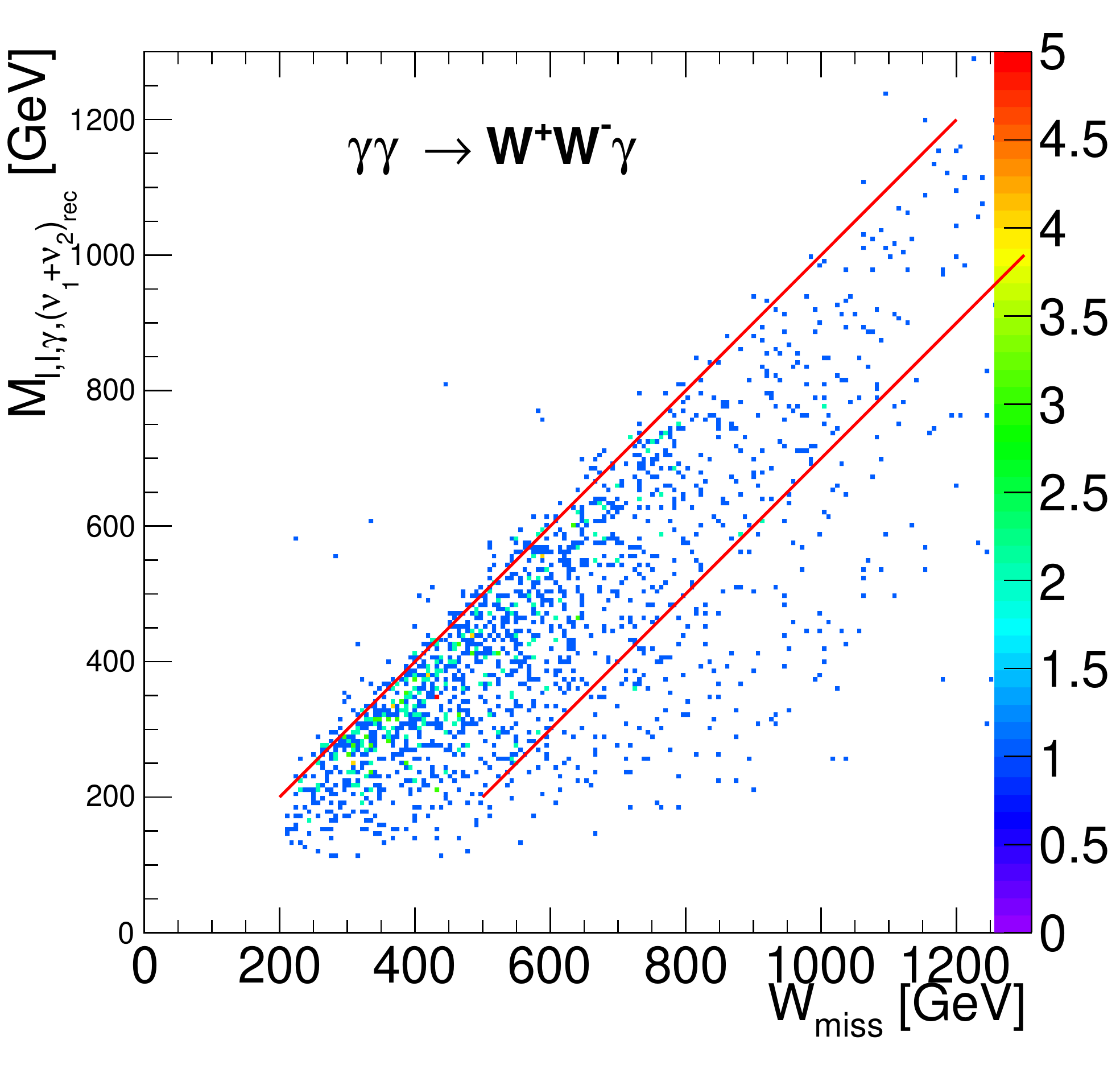}}
          \resizebox{0.45\textwidth}{!}{\includegraphics{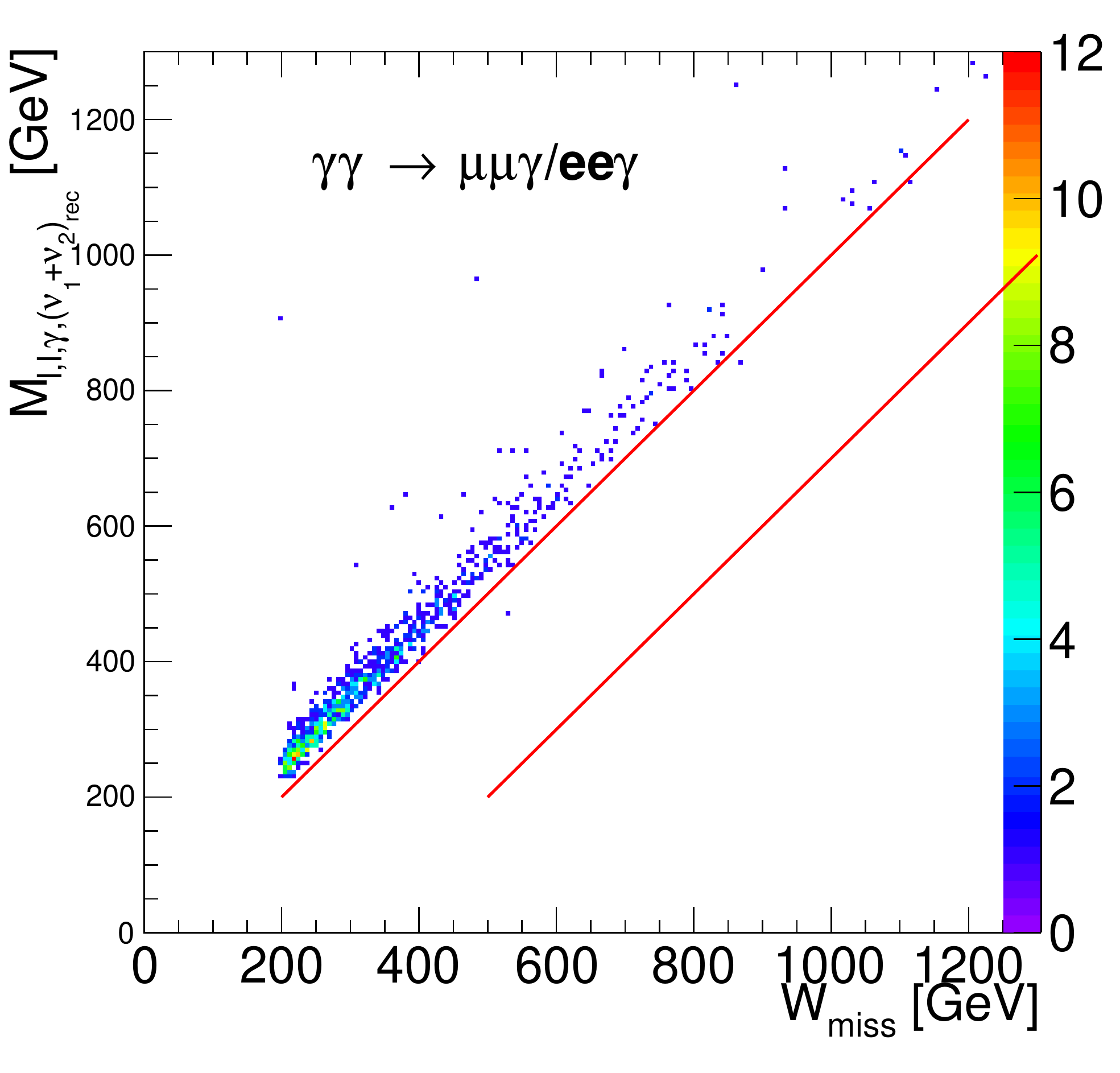}}
          \resizebox{0.45\textwidth}{!}{\includegraphics{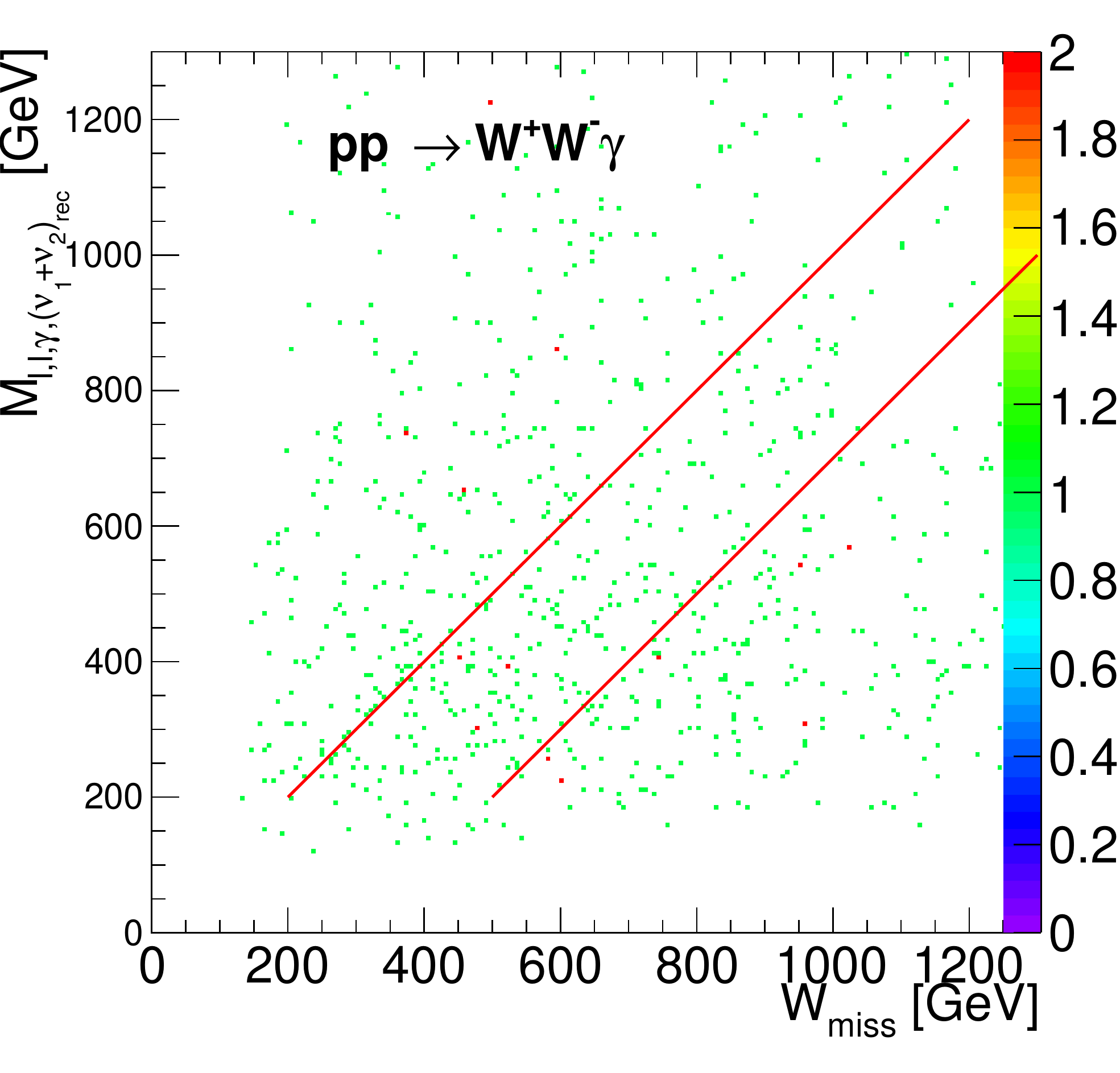}}
          \resizebox{0.45\textwidth}{!}{\includegraphics{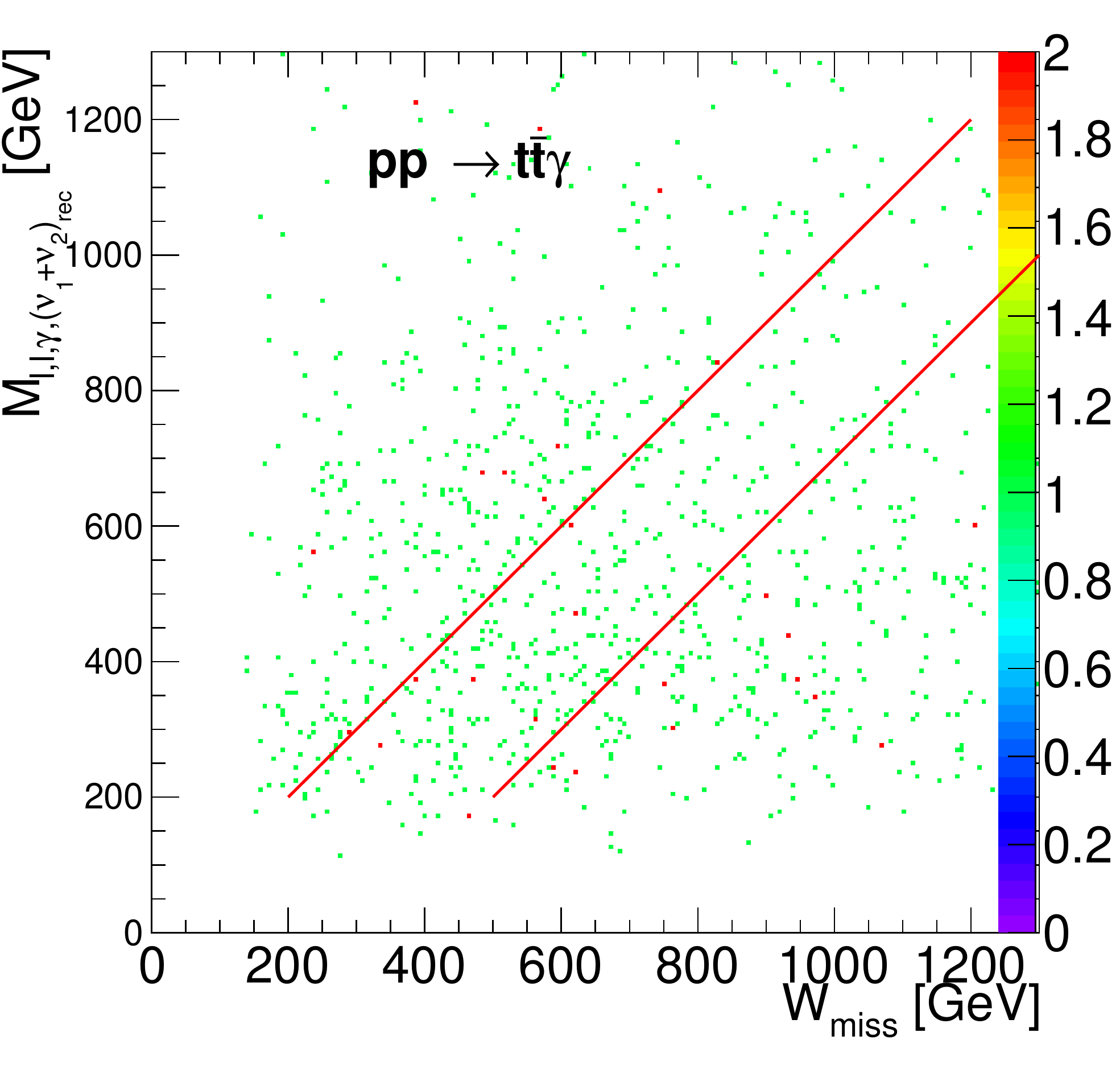}}  
		\caption{Two-dimensional distributions of 
                   reconstructed invariant mass of
                  central system  versus protons
                  missing mass. The constructed invariant mass is obtained by summation of four-momentum
              for two leptons, photon, and four-momentum sum 
          of two neutrinos. The left-top and right-top plots
                  show these two-dimensional distributions for
                  $\gamma\gamma\rightarrow W^+W^-\gamma$  and $
                  \gamma\gamma\rightarrow l^+l^-\gamma$ processes,
                  respectively. The left-bottom and right-bottom plots
                 belong to the inclusive $W^+W^-\gamma$ and
              $t\bar{t}\gamma$ processes superimposed by pile-up protons. The events surrounded between two
              red lines will be kept by the Type III cut.}\label{mreconst}
	\end{center}
\end{figure}
Then, one could obtain the invariant mass of the central system  by summing
over four-momentum of leptons, photons, and~ total missing energy  as $M_{l_1 l_2\gamma (\nu_1+\nu_2)_\text{rec}}=(p({l_1})+p({l_2})+p(\gamma)+p{(\nu_1+\nu_2)_\text{rec}})^2$. However,  we are not fully aware of four-momentum components of each
neutrino that lead to expansion of correlation between the
reconstructed invariant mass of the central system and
$W_\text{miss}$. Figure~\ref{mreconst} top-left indicates this behavior
for the CEP $W^{+}W^{-}\gamma$ process. Having this characteristic, we
require the  0 $ < W_\text{miss}-M_{{l_1 l_2\gamma (\nu_1+\nu_2)}_\text{rec}}<$ 300 which is the region
depicted between the red lines.
Fortunately, even looser correlation is sufficient to
reject inclusive $W^{+}W^{-}\gamma$ ($t\bar{t}\gamma$) backgrounds
showing in the bottom left (right),
respectively. Table~\ref{tab:Cutflow1} shows the number of inclusive
backgrounds after applying each set of selection cuts. The final
yields are represented for 300 $\infb$ IL.   
\begin{table*}[tb]	
		\begin{center} 
			\scalebox{0.83}{
				\makebox[\textwidth][c]{					
					\begin{tabular}{|c|c|c|c|c|c|c|c|}
						\hline
					&	\multirow{2}{*}{ ($\mathcal{L}=300\infb, \sqrt{s}=13\TeV$)}&  $\mathrm{p}\mathrm{p}\rightarrow \tau \bar{\tau} \gamma$  &  $\mathrm{p}\mathrm{p}\rightarrow t \bar{t} \gamma$ &  $\mathrm{p}\mathrm{p}\rightarrow W^+ W^- \gamma$ 	& $\mathrm{p}\mathrm{p}\rightarrow ZZ \gamma$  &$\mathrm{p}\mathrm{p}\rightarrow W^\pm Z \gamma$ & $\mathrm{p}\mathrm{p}\rightarrow l \bar{l} \gamma$ \\ 
					&	& $e\mu (ee+\mu\mu)$ & $e\mu (ee+\mu\mu)$&$e\mu (ee+\mu\mu)$ &$ e\mu (ee+\mu\mu)$& $e\mu (ee+\mu\mu)$& $e\mu (ee+\mu\mu)$ \\
						\hline \hline

 \multirow{5}{*}{Type I } &$p_{T,l_1}>20{\GeV}, p_{T,l_2}>10{\GeV}$  & \multirow{2}{*}{ 8519 (8411)}	&  \multirow{2}{*}{6106(6089)}    & \multirow{2}{*}{ 560 (588)} 	&\multirow{2}{*}{0.07 (4)}  & \multirow{2}{*}{23 (114)}& \multirow{2}{*}{949 (418763)}
						\\  
						&$|\eta_{l_1,l_2}|<2.5$, $iso<0.15$          &&&&&&                     

						\\ \cline{2-8}
						
	&	$\slashed{E}>30{\GeV}$           
  &4106 (4191) & 5447(5409)  & 447 (469) & 0.05 (2.2) &18 (92)&501 (171067)

						\\ \cline{2-8}
						
	&	$p_{T,\gamma}>20{\GeV}$,$|\eta_{\gamma}|<2.5$,  $iso<0.15$                                       &\multirow{2}{*}{ 1141 (1124)  }  &  \multirow{2}{*}{2709(2516)} & \multirow{2}{*}{ 204 (210)}  &\multirow{2}{*}{ 0.02 (1)}  & \multirow{2}{*}{8 (41) }&\multirow{2}{*}{43 (60093)}
						\\  
					&	$\Delta R_{\gamma,l_1}>0.5$, $\Delta R_{\gamma,l_2}>0.5$   &&&&&&                     
							\\ \cline{2-8}
						& Veto $N_j>2$, $p_{T,j}>40{\GeV}$ & 1124(1119) & 1101(1034)&  201(205) &  0.02(0.82) &  7.77(39) & 33(59649)
					
						\\ \cline{2-8}
	&$ |M_{l_1l_2}-m_Z|>10\GeV$ &  1090 (1107) & 282 (266)& 609 (622) &  0.015(0.07) & 6.9 (8.8)& 5(4446)

						\\ \hline
						
  \multirow{5}{*}{Type II } 	&	 $0.008<\xi<0.2$  & 182(187) & 207(197) &  38(39) & 0.005(0.01) &  1.2(1.8) & 0(1035)
						\\  
	&	 $0.008<\xi<0.5$  &  858(772) &   219(202) &  137(140) & 0.01 (0.06)  &  5.3 (7) & 19(3396)                
						\\ \cline{2-8}
&	\multirow{2}{*}{TOF } &0 (0)  & 5.4(3.4) &  0.88(0.75)	& 0(0)& 0.03 (0.006) & 0(19)\\
    	&	& 23(23) &  17(24) & 3.3 (4)& 0(0.0009)& 0.14(0.2) & 0 (100)
                                          \\\hline
 \multirow{2}{*}{Type III } & \multirow{2}{*}{$  W_\text{miss}>200 \GeV$ } &  0(0) &  5.4(2.75) &  0.76 (0.75)
	& 0 (0) &  0.03(0.006) & 0(19)\\  	&&  23(23)	 & 17 (24)  &  3.3(4)	& 0(0.0009)& 0.14(0.2) &0 (100)	\\ \cline{2-8}
                                         
  & \multirow{2}{*}{0 $ < W_\text{miss}-M_{l_1 l_2\gamma (\nu_1+\nu_2)_\text{rec}} <$ 300 } &
   0 (0)  &  0(0)& 0.13 (0.06) & 0 (0)&  0(0) & 0(0) \\  
    	&	&0 (0)  & 1.36(0)  & 0.126 (0.126)	& 0(0.0009)&0.006(0.006) &0(0)	\\ \hline
			\end{tabular}}}
			\vspace{1cm}									
			\caption{The remaining yields of
					the SM inclusive
					background processes coincide with pile-up
					protons 
					after each type of selection cuts for 
					$300\infb$ IL are presented. The mean number of modeled
					pile-up set to 30.}	\label{tab:Cutflow1}
		\end{center}
		
              \end{table*}

	\begin{table*}[tb]	
	\begin{center}
		\scalebox{0.95}{
			\makebox[\textwidth][c]{
				
				\begin{tabular}{|c|c|c|c|c|c|c|}
					\hline
					&	\multirow{2}{*}{($\mathcal{L}=300\infb, \sqrt{s}=13\TeV$)}& $\gamma \gamma\rightarrow  W^+W^-\gamma$ & $\gamma \gamma\rightarrow \tau \bar{\tau} \gamma$  &$ \gamma \gamma \rightarrow l^+l^- \gamma$   \\ 
					&	& $e\mu (ee+\mu\mu)$ & $e\mu (ee+\mu\mu)$&$e\mu (ee+\mu\mu)$  \\
					\hline \hline

					\multirow{5}{*}{Type I } &\small{$p_{T,l_1}>20{\GeV}, p_{T,l_2}>10{\GeV}$}   & \multirow{2}{*}{ 3.3 (3.4)}           
					&  \multirow{2}{*}{2.1 (2.2) }    & \multirow{2}{*}{ 0.7 (464)}  
					
					\\  
					&\small{$|\eta_{l_1,l_2}|<2.5$, $iso<0.15$ }         & & &                         

					\\ \cline{2-5}
					
					&\small{	$\slashed{E}>30{\GeV}$ }             & 2.9 (2.9) &1.3 (1.4) & 0.34 (188)

					\\ \cline{2-5}
					
					&	\small{$p_{T,\gamma}>20{\GeV}$,$|\eta_{\gamma}|<2.5$,$iso<0.15$  }                                      &\multirow{2}{*}{1.5 (1.5) }  &  \multirow{2}{*}{ 0.5 (0.5) }
					& \multirow{2}{*}{ 0.02 (60)} 
					\\  
					&\small{	$\Delta R_{\gamma,l_1}>0.5$, $\Delta R_{\gamma,l_2}>0.5$}      &  & &                       

					\\ \cline{2-5}
					&\small{Veto~$ N_j>2$, $p_{T,j}>40{\GeV}$} & 1.5 (1.5) &0.5 (0.4) & 0.03 (60) 
						\\ \cline{2-5}
					&\small$ |M_{l_1l_2}-m_Z|>10\GeV$ &1.4 (1.4) & 0.46 (0.4)&0.01 (52)
					
					\\ \hline
					
					\multirow{3}{*}{Type II } 	&\multirow{1}{*}{ \small$0.008<\xi<0.2$,TOF } & \multirow{1}{*}{1.2 (1.2)} &\multirow{1}{*}{0.2 (0.3)}   &\multirow{1}{*}{ 0 (17)}
					
					\\  
					&\multirow{1}{*}{\small $0.008<\xi<0.5$, TOF}  & \multirow{1}{*}{1.28 (1.26) } & \multirow{1}{*}{ 0.22 (0.27)}  &\multirow{1}{*}{0 (17)}                       
					
					\\\hline
					\multirow{2}{*}{Type III } & \multirow{2}{*}{\small{$  W_\text{miss}>200 \GeV$ }} & 1.2 (1.2)  & 0.18 (0.2)  & 0 (12) 
					\\ 	& 	& 1.28 (1.26) &0.18 (0.23)  & 0 (12)
					\\ \cline{2-5}
					
					& \multirow{2}{*}{\small{0 $ < 
							W_\text{miss}-M_{l_1 l_2\gamma (\nu_1+\nu_2)_\text{rec}} <$ 300} } &0.94 (0.94) & 0.15 (0.2)  & 0 (0.066)
					\\  	&	& 0.98 (0.99)  & 0.15 (0.2)  & 0 (0.067)
					\\ \hline
		\end{tabular}}}
		\vspace{1cm}									
		\caption{The remaining yields of the SM
                  $\gamma \gamma\rightarrow  W^+W^-\gamma$, $\gamma
                  \gamma\rightarrow \tau^{+}\tau^{-} \gamma$, and $
                  \gamma \gamma\rightarrow e^{+}e^{-}\gamma/\mu^{+}\mu^{-}\gamma$
                  after each type of selection cuts for $300\infb$ IL are presented.}\label{tab:cutflowaa}
	\end{center}
	
\end{table*}

 \subsection{{\boldmath$\gamma\gamma$}-initiated background processes}
One of the main sources of backgrounds to our exclusive $ W^+
W^-\gamma$ process is from the processes with the same mechanism
of photon-photon interactions at the LHC. Among them, the
dominant background is the production of  $\gamma \gamma \rightarrow
l^+ l^-\gamma$ which the leptons can be either electron, muon or tau
(if tau decays leptonically). The $e^{-}e^{+}\gamma,~\mu^{-}\mu^{+}\gamma$ channels
are dominant in the SF signal region while the $\tau^{-}\tau^{+}\gamma$
can contribute equally in both SF and DF signal
regions. We applied the three types of selection cuts described in the
previous sections. Table~\ref{tab:cutflowaa} shows the number of events
for the SM $W^+ W^- \gamma$ exclusive process and their photon-photon
initiated backgrounds after each set of selection cuts and assuming
300 $\infb$ expected IL. It is interesting to mention that the
correlation between the reconstructed mass of the central system and
protons missing mass for $l^+l^-\gamma$ background which is depicted
in  Figure~\ref{mreconst} (top-right) is completely different
from the SM $W^+ W^- \gamma$  mass correlation. Table~\ref{tab:cutflowaa} shows a large amount
of this background remains after type I and II, but reduced to the zero level considering this cut.

 \subsection{Double pomeron exchange processes}
In addition to central exclusive production via $\gamma\gamma$
interaction, $W^+ W^- \gamma$  and $l^+ l^- \gamma$
processes can occur through the double
pomeron exchange (DPE). The pomeron is believed to carry the quantum
numbers of vacuum, thus they will be colorless states in QCD
language. It is also proposed that pomeron has partonic structure
such as hadrons \cite{Pomeron:1, Nussinov:1975qb}. Therefore, hard diffractive processes can be described in
terms of single and double pomeron exchange between two
protons based on the Ingelman-Schlein approach
\cite{Ingelman:1984ns}  that has been searched in the different
	experiments ever since \cite{Bonino:1988ae,Ahmed:1995ns, Abe:1997rg}. In this
model, cross section of the DPE can be factorized into the diffractive
parton distribution functions and matrix element of
hard interaction between the pomeron constituents that considered to
be gluonic. Currently, several MC generators such as  Forward Physics Monte Carlo (FPMC) generator \cite{Boonekamp:2011ky} can calculate the cross
section and generate events of the DPE processes such as dilepton, di-boson, and
di-jet productions. Since our favorite
$\text{DPE} \rightarrow W^+ W^- \gamma$  and $\text{DPE}
\rightarrow l^+ l^- \gamma$ processes are not yet implemented in any generators,  we inevitably have considered some assumptions in order
to extract their contribution into the SM signal and
background processes. According to \cite{Ingelman:1984ns} the cross section of $\text{DPE} \rightarrow W^+
W^- \gamma$ and $\text{DPE}
\rightarrow l^+ l^- \gamma$ can be factorized into the scattering
amplitude of emerging partons from each pomeron
and already known diffractive PDF. On the other hand the cross section
of inclusive $W^+ W^- \gamma$  and $l^+ l^- \gamma$ from proton-proton collision
also can be factorized into the matrix element of partonic
interaction by the PDF of protons. Therefore, one can assume the
equality of cross section ratios of inclusive and DPE as following

\begin{equation}
  \frac{\sigma_{\mathrm{p}\mathrm{p}\rightarrow W^{+}W^{-}/l^+l^-}}{\sigma_{\mathrm{p}\mathrm{p}\rightarrow
W^{+}W^{-}\gamma /l^+l^-\gamma}}=\frac{\sigma_{\text{DPE}\rightarrow W^{+}W^{-}/l^+l^-}}{\sigma_{\text{DPE}\rightarrow
W^{+}W^{-}\gamma /l^+l^-\gamma}}. \label{ratio}
\end{equation}

We calculated the ratios $\frac{\sigma_{\mathrm{p}\mathrm{p}\rightarrow W^{+}W^{-}/l^+l^-}}{\sigma_{\mathrm{p}\mathrm{p}\rightarrow
W^{+}W^{-}\gamma/l^+l^-\gamma}}$  by \texttt{MadGraph5\_aMC@NLO} at
tree level for $W^+ W^- \gamma$  and $l^+ l^- \gamma$ processes
resulting to values of 243.3 and 92.5, respectively. Due to the
existence of a photon in the matrix element of the denominator, one would expect
the ratio to be around $1/\alpha\simeq$ 100 where the $\alpha$ is
Fine-structure constant. The  di-lepton production to the
di-lepton plus photon cross section ratio seems to agree with this expectation, but the
di-boson  ratio is more than twice as high the expected ratio. The reason is behind the contributed Feynman diagrams with triple and quartic gauge
boson couplings which have destructive interference and lead to
lower cross section ratio w.r.t di-leptonic ratio. We also calculate
the cross sections of DPE$\rightarrow
W^{+}W^{-}$ and DPE$\rightarrow
l^{+}l^{-}$ using FPMC generator at 13 TeV and obtain the corresponding values of 
1.35 and 701.4 pb for these two processes, respectively. Having the left-hand-side of equation
\ref{ratio} also the numerator of right-hand-side we obtain 5.5 fb and 7.58 pb
for the cross section of DPE$\rightarrow
W^{+}W^{-}\gamma$ and DPE$\rightarrow
l^{+}l^{-}\gamma$, respectively. Furthermore, the obtained cross sections have to be
multiplied by a gap survival probability for QCD diffractive and
central exclusive productions which is 0.03\cite{Martin:2008nx}. This
factor accounts for the probability that the gaps are surviving from 
the presence of extra particles in the interaction. This rapidity gap can be 
washed out mainly by soft inelastic interaction that produces some
secondary particles or re-scattering of leading hadron or hard QCD
bremsstrahlung. In order to estimate the contribution of these two DPE
processes in our signal regions we assumed that the kinematics of
their final state are similar to the $\gamma \gamma \rightarrow
l^+ l^-\gamma$ and $\gamma \gamma \rightarrow
W^+ W^-\gamma$. Therefore, we obtain the efficiency of type I and type
II selection cuts from photon initiated processes and apply
these efficiencies as a new factor to the similar DPE processes. The
summary of factors which are applied to the DPE processes and their
final contributions in the two signal regions using $300 \infb$
IL is shown in Table \ref{dpe}.
\section{ SM $\gamma \gamma \rightarrow W^{+}W^{-}\gamma$ measurement  }\label{sec:smsens}
\begin{figure}
	\begin{center}
          \resizebox{0.45\textwidth}{!}{\includegraphics{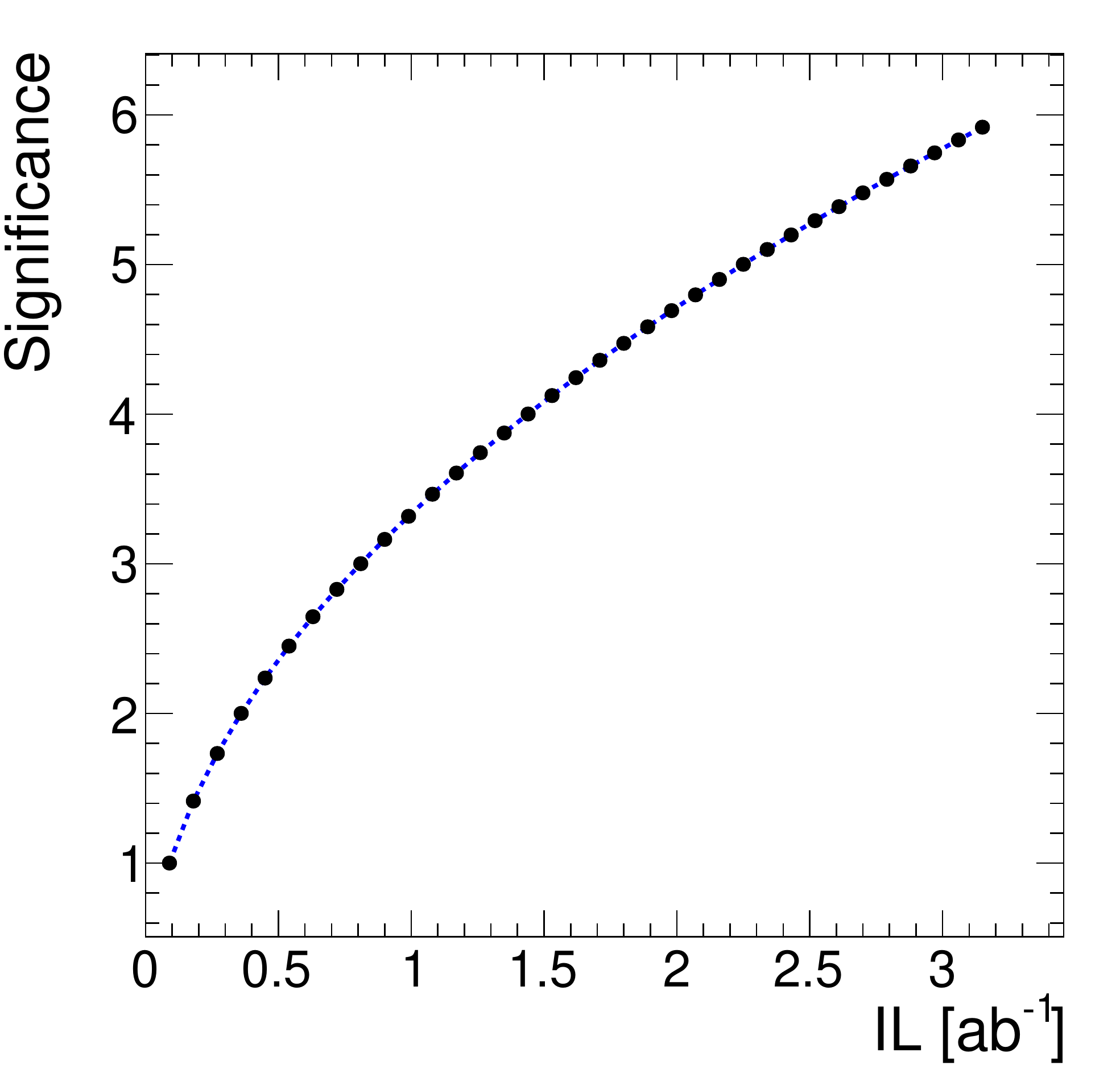}}

		\caption{ Expected significance as a function of
                  IL for the SM central exclusive
                  production of $W^+W^-\gamma$ process using only
                  di-leptonic channel.}\label{sign}
	\end{center}
\end{figure}

Considering the small predicted cross section of central exclusive
$W^{+}W^{-}\gamma$ process, the measurements of this process needs a
high amount of data. On the other hand, the advantage of having timing
and tracking FDs allows us to measure this process in the high pile-up
run conditions of the LHC, also bring the backgrounds of this process in very small amount  as shown in detail in the previous sections.
 We consider the fully leptonic decay of W bosons in the SF and
DF channels. In order to calculate the potential discovery of this
process, we use Profile Likelihood formalism.  The
median significance assuming the signal hypothesis $\mu$=1 can be
obtained by
\begin{equation}
  \text{med}[Z_{0}|1]=\sqrt{2((s+b)ln(1+s/b)-s)},
  \label{sig}
\end{equation}
where s and b are the number of signal and
backgrounds~\cite{Cowan:2010js}. Figure~\ref{sign} illustrates the potential observation of this process considering only di-leptonic channel as a function of IL.

\begin{table}
	\begin{center}
		\begin{tabular}{|l|l|l|}
			\hline
			\multicolumn{3}{|c|}{Backgrounds} \\
			\hline
			\hline
			Process & DPE$\rightarrow W^{+}W^{-}\gamma$&
			DPE$\rightarrow l^{+}l^{-}\gamma$\\ \hline
			Total cross section [fb] & 5.5 &  7583 \\ \hline
			Gap survival rapidity [fb]& 0.165& 227.49 \\ \hline
			Type I selection cut eff [fb]& 0.002  & 0.36 \\ \hline
			Type II selection cut eff [fb]& 4e-6 (1e-4)   & 0e-5 (0e-5) \\ \hline
			Type III selection cut eff [fb]& 0e-6 (0e-4) & 0e-5 (0e-5) \\ \hline
			Final yield  for 300 $\infb~(ee,\mu\mu,e\mu))$& 0e-6 (0e-4)  & 0e-5 (0e-5) \\ \hline
		\end{tabular}
	\end{center}
	\caption{The sequence of different type of cuts on the cross
		sections of DPE$\rightarrow
		W^{+}W^{-}\gamma$ and DPE$\rightarrow
		l^{+}l^{-}\gamma$  processes in two signal regions are presented. The last row represents the final yields for 300 $\infb $ IL.} \label{dpe}
\end{table}
The amount of data for having a strong evidence of this process with
3$\sigma$ significance is about the 0.8 $\inab$ while for full
observation of this process with 5$\sigma$ significance one expected
2.1 $\inab$ IL. The observation of this process is not the most
interesting aspect of this study but rather the power of this process
to constrain anomalous couplings due to very low amount of backgrounds
 which is going to be discussed in next two
 sections. It should be mentioned that estimated
   amount of data needed to observe this process, given in
   Figure~\ref{sign}, is obtained from the
   extrapolation of the present study based on
   $<N_{PU}>=30$.

\section{Sensitivity to anomalous gauge boson couplings} \label{sec:limits}

\begin{figure}
	\begin{center}
		\vspace{1cm}
		\resizebox{0.45\textwidth}{!}{\includegraphics{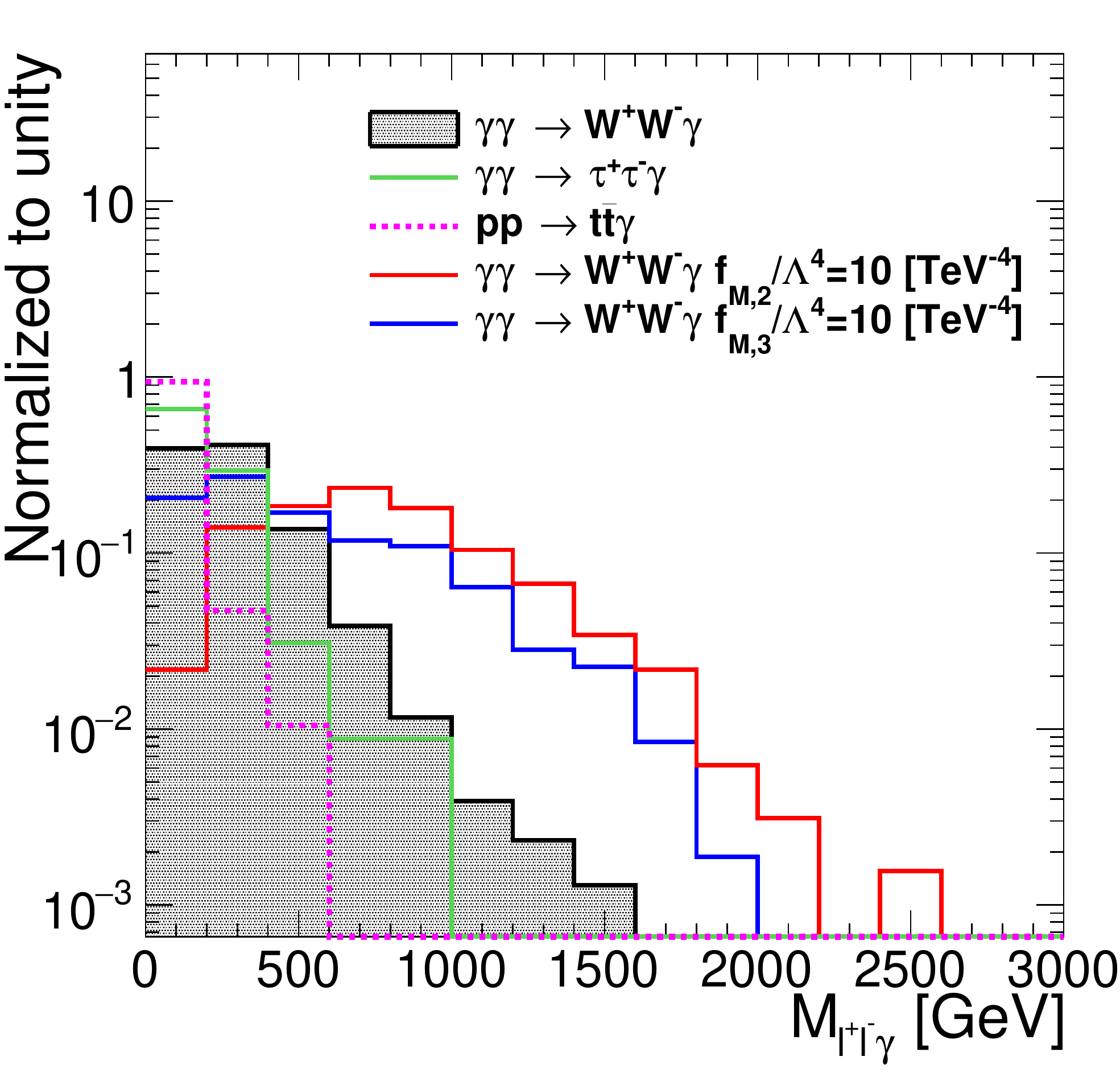}}  
		
		\caption{The invariant mass distribution of
                  $l_1l_2\gamma$,  measured in the central detector. The
                  SM backgrounds are indicated in shaded gray, solid
                  green and dashed purple
                  histograms. The red and blue histograms show the two
                  samples of dimension-8 coupling $f_{M,2}/\Lambda^4=f_{M,3}/\Lambda^4=10\TeV^{-4}$. }\label{mllgam}
	\end{center}
      \end{figure}

In this section, we discuss the potential of $W^{+}W^{-}\gamma$ CEP to
probe aTGCs and aQGCs at the LHC using forward
detectors. In order to reach the highest sensitivity, one needs to
study these new couplings in the signal dominated region.  Therefore,
we need to modify the type III of introduced cuts in the previous
sections. The first modification is to  restrict the lower cut on the protons missing mass to  more than 900 GeV, as the contribution of
anomalous couplings is enhanced at high missing mass values while the
backgrounds are effectively suppressed.  Figure~\ref{kisilostmass} right shows the
distribution of protons missing mass for two scenarios of anomalous
couplings, SM $W^{+}W^{-}\gamma$ CEP as the irreducible background,
and some other photon initiated backgrounds. In addition to this
change, we introduce a new cut on the invariant mass of visible central
state $\ie$ lepton pair and photon. We
restrict the invariant mass of lepton pair and photon to be higher
than 200 and 500 GeV for the two considered scenarios of
acceptance. Even though this cut is highly correlated with the
previous cuts for photon initiated backgrounds, it is fully
independent for inclusive backgrounds with pile-up protons. Therefore, this cut effectively suppresses inclusive backgrounds while it is safe for keeping the signal
contribution. Figure~\ref {mllgam} illustrates the invariant mass of
lepton pair and photon for some of photon initiated and inclusive
backgrounds as well as two considered anomalous signals. We also
require the difference between the missing mass of protons and
reconstructed mass state only to be greater than zero as we observed the
correlation between these two masses behave differently from what
indicated in Figure~\ref{mreconst} top-left when one includes the
anomalous couplings. Thus, to avoid the drop in signal efficiency we
loosen this criterion.

\begin{table}
  	\begin{center}
  		\hspace{-1cm}
  			\scalebox{0.8}{
	\begin{tabular}{|c|c|c|c|c|c|}
					\hline
						\multirow{2}{*}{\small{($\mathcal{L}=300\infb, \sqrt{s}=13\TeV$)}}& \small{Backgrounds}& \small{$\lambda=0.05$} &\small{$f_{M,1}/\Lambda^4=10\TeV^{-4}$}&\small{$f_{M,3}/\Lambda^4=10\TeV^{-4}$}  \\ 
					&	 $e\mu+ee+\mu\mu$ &$e\mu+ee+\mu\mu$&$e\mu+ee+\mu\mu$&$e\mu+ee+\mu\mu$\\	\hline 
				{Type I }        &6745.9 & 116 &3.5 & 64	\\ \hline
					                                      \small{TOF,$ 0.008<\xi< 0.2 (0.5)$} & 38.8 (71.3)& 7(85) &1.8 (2.9) & 3 (43) \\\hline	
			\small{ $  W_\text{miss}>900,
          M_{l^{+}l^{-}\gamma}>200 (500) \GeV,$} & \multirow{2}{*}{0.3 (0.9)}  & \multirow{2}{*}{ 7(79)} &\multirow{2}{*}{0.3 (1.1)  }&\multirow{2}{*}{2 (38)} \\
          $ W_\text{miss}-M_{l_1 l_2\gamma (\nu_{1}+\nu_{2})_\text{rec}} >0 $&&&&	\\ \hline		
		\end{tabular}}									
      \end{center}
      	\caption{The yield of the SM backgrounds and few
          signal samples after applying the Type I cut, modified
          Type II, and Type III cuts for 300 f$\text{b}^{-1}$
          IL. The considered signal samples are  $\lambda=0.05$ and  $f_{M,1}/\Lambda^4=f_{M,3}/\Lambda^4=10\TeV^{-4}$.} \label{aqgcflow}
              \end{table}

              Table~\ref{aqgcflow} shows the effect of each type of
              selection cuts on the total SM backgrounds including the
              photon initiated, DPE, and inclusive backgrounds and few
              cases of aQGCs and aTGCs.

              An important subject in studying the anomalous gauge boson
              couplings is to check for the preservation of unitarity. It is well
              understood that the non-zero value of new
              EFT operators could result in the rapid increase of
              scattering amplitude w.r.t the energy which could
              violate the unitarity in the sufficient high center-of-mass energy
              of colliding partons. In this analysis, the presence of
              dimension-6 and -8 effective operators could potentially cause this violation. However, considering the
              FDs acceptance e.g 0.008$< \xi <$ 0.2 prevents the
              center-of-mass energy of two colliding photons from exceeding  2.5 GeV
               which is shown to be approximately
              safe~\cite{Corbett:2014ora}. For other cases in which
              the acceptance cut is not sufficient to exclude the
              regions which unitarity is not preserved one usually uses
              the form factors (FFs). These FFs essentially are energy
              dependent cutoff of a complete model at the scale of $\Lambda$ which is integrated
              out as the higher-order EFT operators. A dipole FF or a sharp cutoff on the EFT operators at a
              fixed energy scale is usually considered to
              control the unitarity. In the EFT description
              which is a model-independent approach, there is no preferred method
              or functionality for FFs. Therefore, various
              forms of FFs are considered in the literature. In this analysis, to compare our results in a FF independent way with several
              experimental measurements~\cite{{Khachatryan:2016vif:wgamma8tevcms} , {Khachatryan:zga8cms}, {Aaboud:zga8atlas},{Khachatryan:2016mud:exclwwcms}, {Sirunyan:wzv13cms}}, we also do not apply
              any unitarity dipole form factor or cutoff.
             
\subsection{Statistical method} \label{sec:statmethod}

In this section, we discuss the potential of this channel to
constrain aTGCs and
aQGCs assuming  one and two-dimensional scan of
effective couplings. We use the signal region defined in 
Table~\ref{aqgcflow} in order to count the contribution of signal and SM
backgrounds in both  SF and DF di-leptonic channels. In order to
extract the two-dimensional expected limit on a pair of effective couplings, we define the profile likelihood test
statistics as follows

\begin{equation}
  q(c_{i}^{d},c_{j}^{d})=-2\text{ln} \frac{\mathcal{L}(n|f(c_{i},c_{j})+b, \hat{\hat{\theta}})}{\mathcal{L}(n|f(\hat{c_{i}},\hat{c_{j}})+b, \hat{\theta})}.
  \label{eq:cep}
 \end{equation}
 The $\mathcal{L}$ is the product of Poisson distribution of
 expected events and log normal distribution of nuisance
 parameters denoted by $\theta$ for each dileptonic signal
 region. $\hat{c_{i,j}}$ and $\hat{\theta}$ are the values of
 parameter of interest
 and nuisance parameters which
 maximize the likelihood. The $\hat{\hat{\theta}}$ is the value of
 nuisance parameters that maximize the likelihood of the couplings
 which are being tested. $f(c_{i},c_{j})$ is the yield of 
 anomalous couplings (ACs) plus SM $W^{+}W^{-}\gamma$ for a 
 specific pair of couplings and b is the number of other SM
 backgrounds.
 In order to scan the test statistics over the different values of ACs,
 one needs to have SM+AC yield as a function of ACs. To obtain the functionality, we  generate signal sample while switching on the two
 couplings simultaneously using the re-weighting method in
 \texttt{MadGraph5\_aMC@NLO} for more than
 100 different sets of couplings. Then the yield functionality
 obtained by fitting these 100 points by a Quadratic Polynomial. In
 order to be conservative, we assume 100$\%$ uncertainty on the
 background yields.
 It has been shown that the distribution of defined test statistic
approaches the $\chi^{2}$ distribution~\cite{Wilks:1938dza}. Thus, one
can extract the expected limit by defining the delta log-likelihood
(deltaLL) function. Consequently, 68(95)$\%$ C.L. allowed region of
a pair of parameters can be calculated from $q(c_{i},c_{j})$= 2.30
(5.99).   The same procedure is followed for obtaining the expected limit on one coupling except for the quantile for 68(95) $\%$ C.L. which are $q(c_{i})$= 1.00 (3.84).
\subsection{Triple gauge boson couplings} \label{sec:model1}
\begin{figure}
	\begin{center}
		\vspace{1cm}
		\resizebox{0.45\textwidth}{!}{\includegraphics{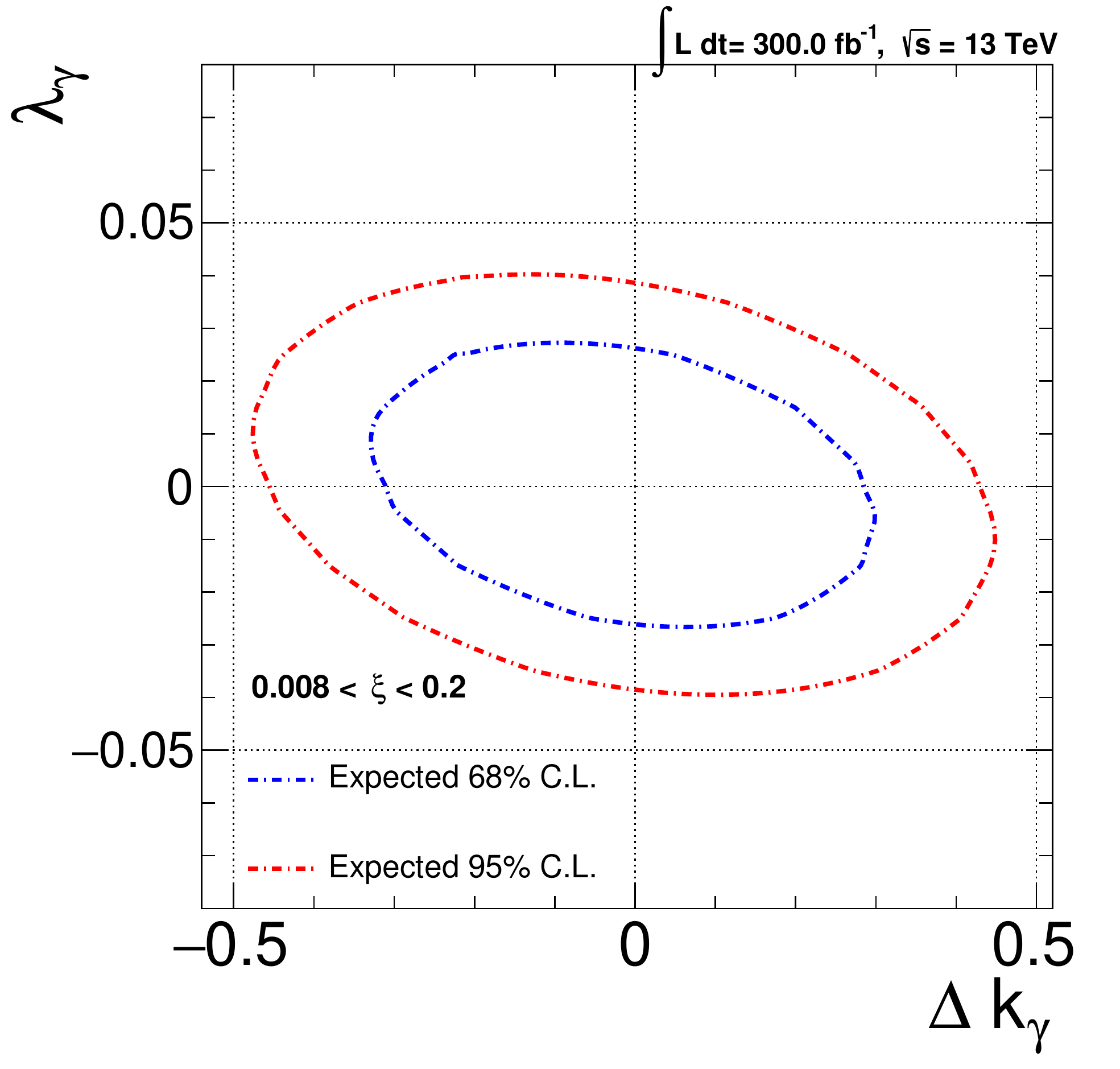}}  
		\resizebox{0.45\textwidth}{!}{\includegraphics{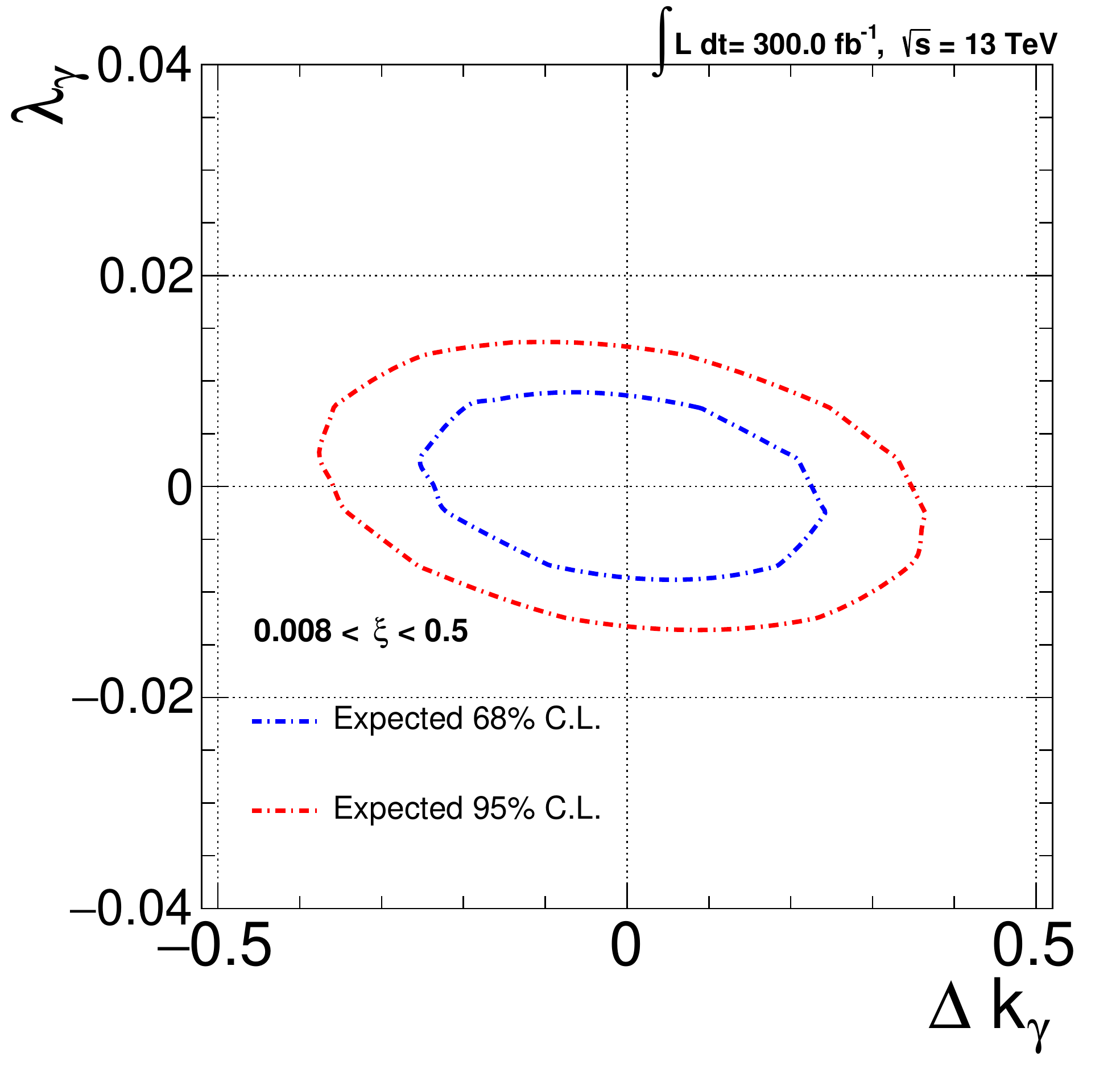}}  
		
		\caption{ Two-dimensional 68$\%$ and 95$\%$ C.L. expected limit on the $\lambda_{\gamma}$ and $\Delta
k_{\gamma}$ couplings assuming 0.008$
< \xi < $0.2 (left) and  0.008$ < \xi < $0.5 (right) and considering
the IL corresponds to 300$\infb$. }\label{2dtgbq:limit}
	\end{center}
\end{figure}

In this section, we calculate the two-dimensional limit on
  $\lambda_{\gamma}$ and $\Delta \kappa_{\gamma}$  as well as one-dimensional  constraints on one of aTGC by setting the 
other one to zero. In this respect, we consider the signal region explained in
the previous part and  summarized  in  Table~\ref{aqgcflow} in order
to select signal events and employ the statistical method
discussed in Section~\ref{sec:statmethod}. Figure~\ref{2dtgbq:limit} indicates
68$\%$ and 95$\%$ C.L expected limit on the $\lambda_{\gamma}$ and $\Delta
k_{\gamma}$ couplings assuming two different acceptance regions,  0.008$
< \xi < $0.2 (left) and  0.008$ < \xi < $0.5 (right) and considering
the IL corresponds to 300$\infb$.

It is obvious that in the higher acceptance regions the sensitivity of the process to the
anomalous parameters especially $\lambda_{\gamma}$ improve as the
main contribution of the signal from this parameter appears at the
high proton missing mass region. The one-dimensional 68$\%$ and 95$\%$
C.L. expected limits on these parameters are presented in 
Table~\ref{table:tgblimit} for both acceptance scenarios.  It should be
mentioned that since the $\Delta \kappa_{\gamma}$ contribution to the 
$W^{+}W^{-}\gamma$ CEP 
only increases the normalization of the SM, therefore, expected
sensitivity of this parameter can be improved by lowering the mass cut
criterion in Table~\ref{aqgcflow}. However, as in general the
sensitivity of this process to the coupling$\Delta \kappa_{\gamma}$ is not  high  enough we
decide to keep the same signal region for both couplings.


\begin{table}
  	\begin{center}
  		\begin{tabular}{|l|l|l|}
  			\hline
  			\multicolumn{3}{|c|}{ ($\mathcal{L}=300\infb, \sqrt{s}=13\TeV$)} \\
  			\hline
  			aTGCs & 0.008$ < \xi < $0.2& 0.008$ < \xi < $0.5 \\ \hline
                                                                      
                 \multirow{2}{*}{$\lambda_{\gamma}$} & 68$\%$ C.L.  [-0.019,0.019] & 68$\%$ C.L.  [-0.006,0.006] \\ 
                      &  95$\%$ C.L.  [-0.032,0.032]   &  95$\%$ C.L.  [-0.011,0.011,]  \\\hline
                  \multirow{2}{*}{$\Delta \kappa_{\gamma}$}&68$\%$ C.L. [-0.16,0.15] & 68$\%$ C.L. [-0.17,0.16] \\ 
                   &95$\%$ C.L. [-0.26,0.25] & 95$\%$ C.L. [0.30,0.29] \\ \hline

  		\end{tabular}
  	\end{center}
  	\caption{68$\%$ and 95$\%$ C.L expected limit on the $\lambda_{\gamma}$ and $\Delta
k_{\gamma}$ couplings assuming two different acceptances,  0.008$
< \xi < $0.2  and  0.008$ < \xi < $0.5, considering
the IL corresponds to 300$\infb$.} \label{table:tgblimit}
  \end{table}

\subsection{ Quartic gauge boson couplings} \label{sec:model2}

\begin{figure}
	\begin{center}
		\vspace{1cm}
		\resizebox{0.45\textwidth}{!}{\includegraphics{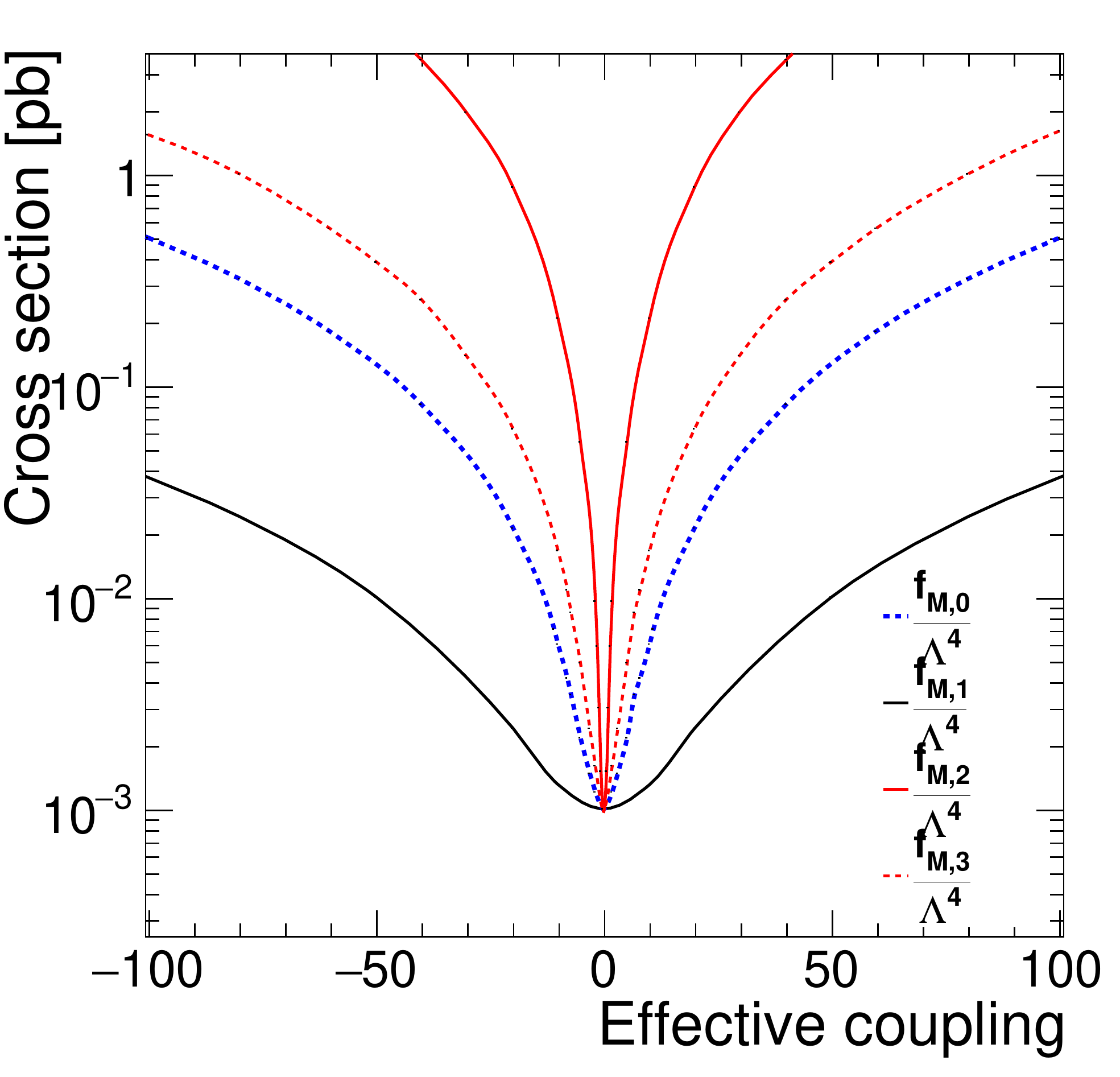}}  
		
		\caption{The cross sections of $\mathrm{p}\mathrm{p}\rightarrow
                  \mathrm{p}W^{+}W^{-}\gamma \mathrm{p}$ as a function of four
                  dimension-8 anomalous
                  QGCs at proton-proton center-of-mass energy of 13 TeV.}\label{xsecdim8}
	\end{center}
\end{figure}

In this section, we examine the sensitivity of this process to the
aQGCs arising from dimension-8 effective
operators. It has been shown that the dimension-6 operators could
contribute to both triple and quartic gauge boson
couplings~\cite{Degrande:2013rea}. Therefore, the lowest order of operators which only appear
as quartic couplings is dimension-8.  Figure~\ref{xsecdim8} shows the
cross sections of $W^{+}W^{-}\gamma$ CEP as a function of four main new
quartic couplings which give the contribution to this process via effective
$WW\gamma\gamma$ vertex. 

\begin{figure}
	\begin{center}
		\vspace{1cm}
                \resizebox{0.45\textwidth}{!}{\includegraphics{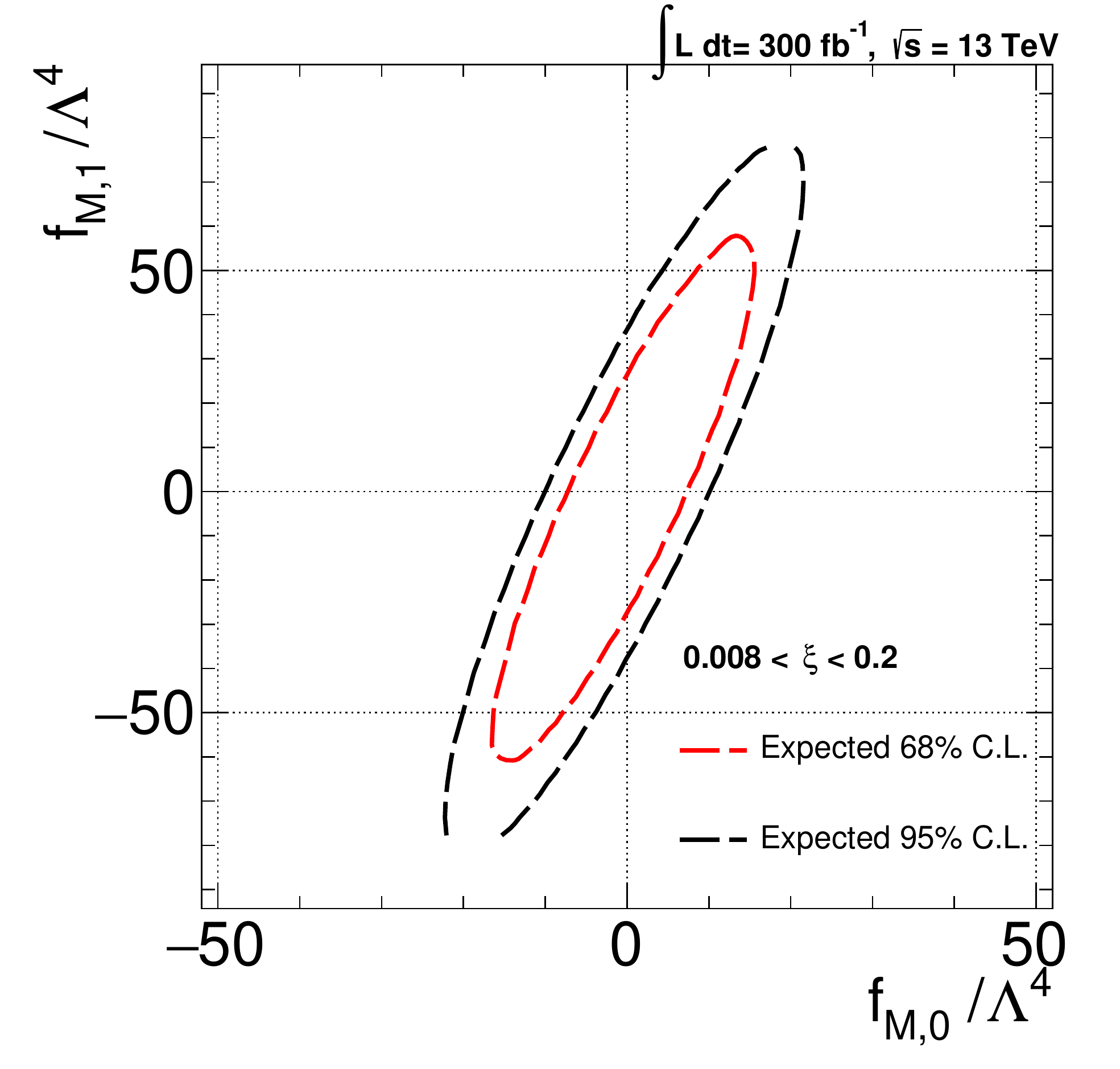}}
                		\resizebox{0.45\textwidth}{!}{\includegraphics{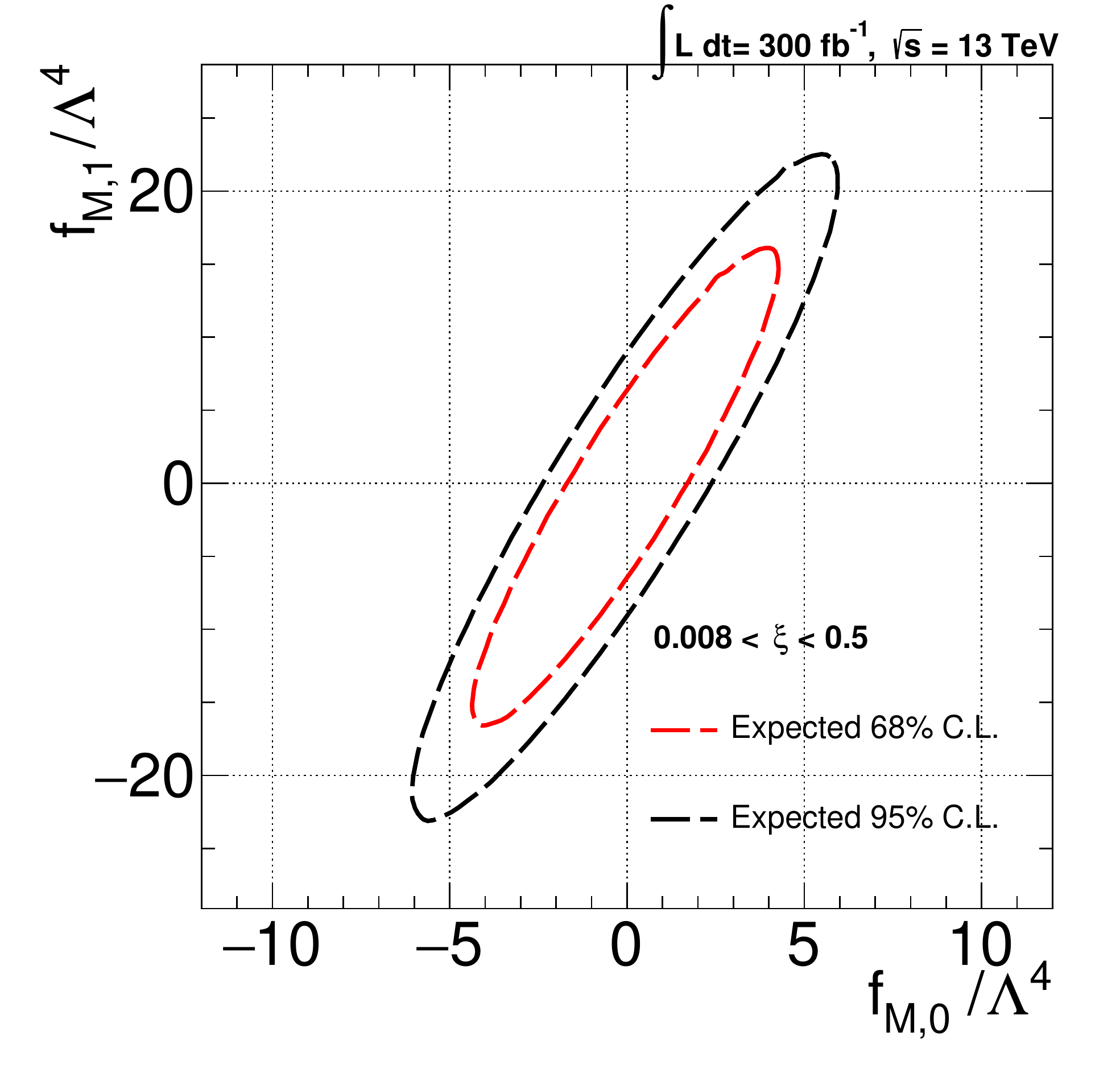}}  
				\resizebox{0.45\textwidth}{!}{\includegraphics{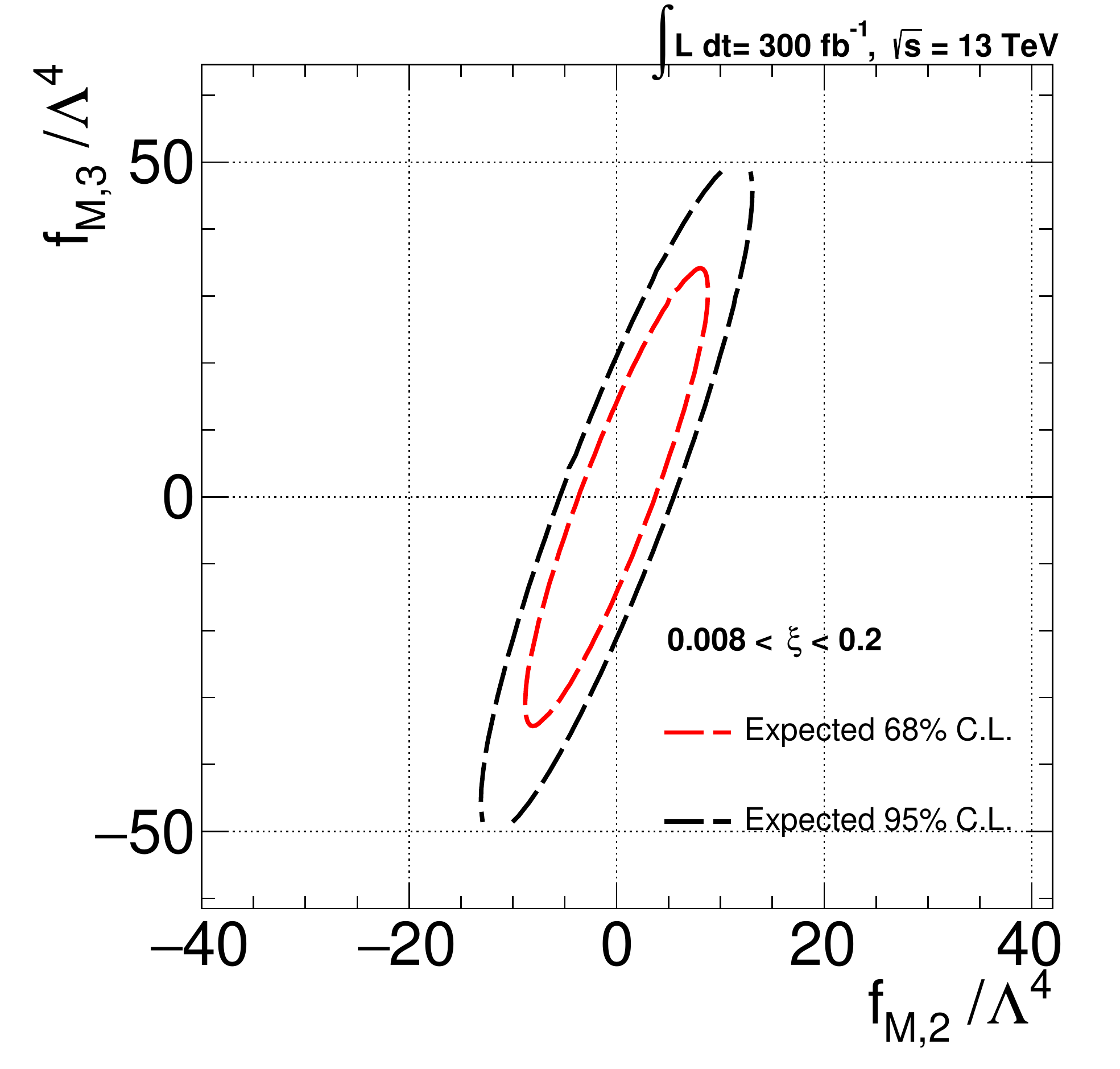}}  
				\resizebox{0.45\textwidth}{!}{\includegraphics{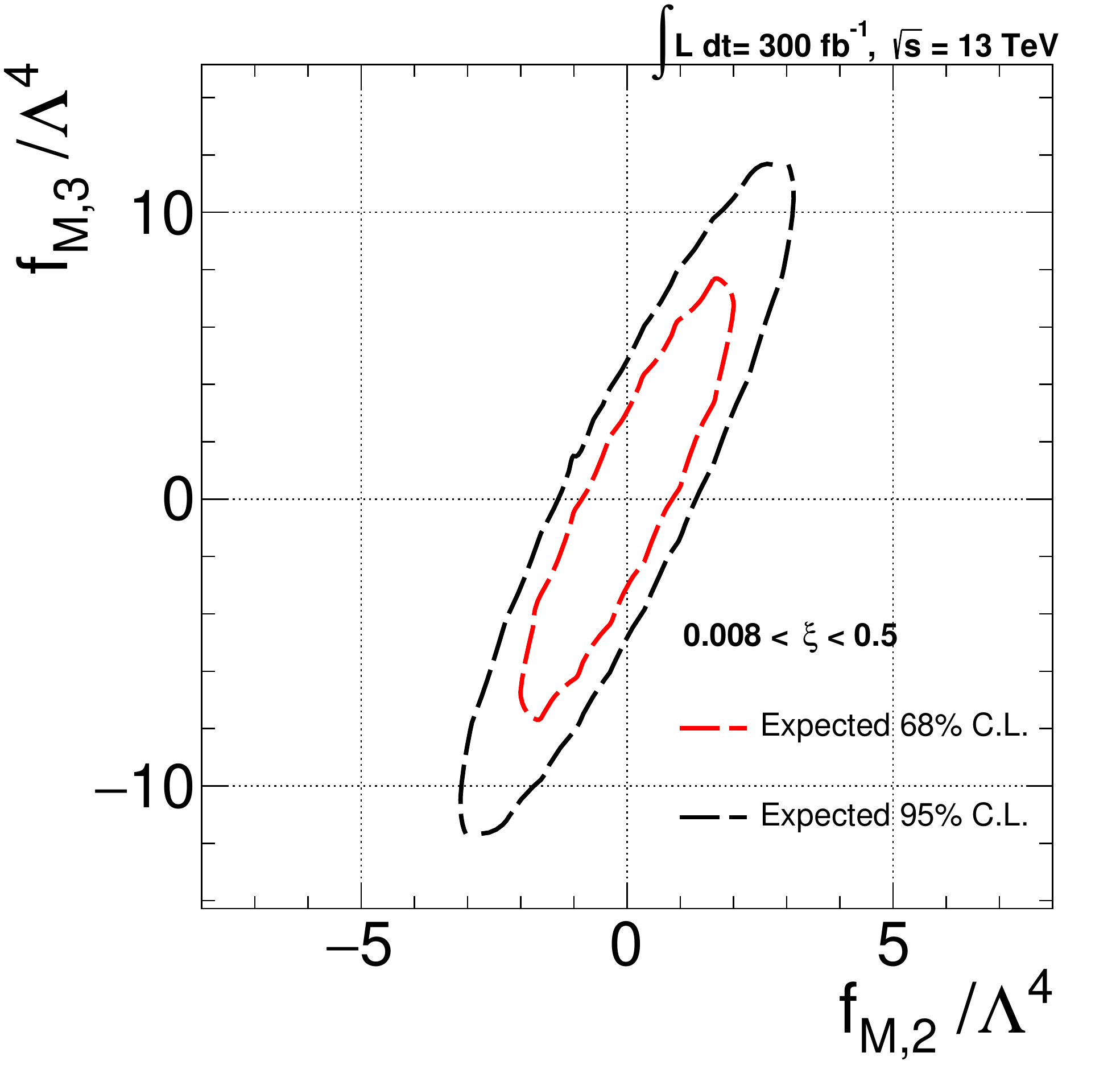}}  

		\caption{Two-dimensional 68$\%$ and 95$\%$ C.L
                  expected allowed regions between $f_{M,1}$ and $f_{M,0}$ for
acceptance of 0.008$ < \xi < $0.2 (top-left) and 0.008$ < \xi < $0.5
(top-right). Also Two-dimensional allowed region for $f_{M,3}$ and
$f_{M,2}$ for acceptance of 0.008$ < \xi < $0.2 (bottom-left) and 0.008$ < \xi < $0.5
(bottom-right) assuming 300 f$\text{b}^{-1}$ IL.}\label{fig:2dtgbqlimit}
	\end{center}
\end{figure}

As it can be seen $f_{M,2}$ and $f_{M,3}$
have strong dependency to the cross section and are expected to give
tighter constraints comparing to the $f_{M,0}$ and $f_{M,1}$. Again in this
section, we used the signal regions defined in 
Table~\ref{aqgcflow}. In the first step, we calculated the expected 68$\%$ and
95$\%$ C.L exclusion regions between two couplings for both
acceptance scenarios. Figure~\ref{fig:2dtgbqlimit} depicts these two
dimensional allowed regions between $f_{M,1}$ and $f_{M,0}$ for
acceptance of 0.008$ < \xi < $0.2 (top-left) and 0.008$ < \xi < $0.5
(top-right). The similar expected exclusion regions between $f_{M,3}$ and
$f_{M,2}$ couplings for 0.008$ < \xi < $0.2 (bottom-left) and 0.008$ < \xi < $0.5
(bottom-right) are shown in Figure~\ref{fig:2dtgbqlimit}.

 We also estimate one-dimensional 68$\%$ and
95$\%$ C.L.  expected  limits on these anomalous
couplings by assuming one coupling as free parameter  and the  rest  couplings are set to zero. The one-dimensional constraints on aQGCs  are presented in Table~\ref{table:quarticlim}. The
one-dimensional reported limit values in Table~\ref{table:quarticlim} include both
acceptance scenarios, also considering
the 300 $\text{fb}^{-1}$ IL.

\begin{table}
  	\begin{center}
  		\begin{tabular}{|l|l|l|}
  			\hline
  			\multicolumn{3}{|c|}{68$\%$ and 95$\%$ Expected limit, ($\mathcal{L}=300\infb, \sqrt{s}=13\TeV$)} \\
  			\hline
  			dimension-8 aQGC & 0.008$ < \xi < $0.2& 0.008$ < \xi < $0.5 \\ \hline
                                                                      
                 \multirow{2}{*}{ $f_{M,0}/\Lambda^4 (\TeV^{-4})$}&68$\%$ C.L. [-5.7,5.7] &  68$\%$ C.L.[-1.3,1.3] \\ 
                   &  95$\%$ C.L. [-8.7,8.7]   &   95$\%$ C.L. [-2.0,2.0] \\ \hline
                  \multirow{2}{*}{ $f_{M,1}/\Lambda^4 (\TeV^{-4})$}& 68$\%$ C.L. [-21.9,21.9] & 68$\%$ C.L. [-5.0,5.0] \\ 
                       &95$\%$ C.L. [-32.8,32.8]   &  95$\%$ C.L.  [-7.7,7.7] \\ \hline
                  \multirow{2}{*}{$f_{M,2}/\Lambda^4 (\TeV^{-4})$}&68$\%$ C.L. [-1.9,1.9] & 68$\%$ C.L. [-0.5,0.5] \\ 
                  &    95$\%$ C.L.  [-3.2,3.2]          &  95$\%$ C.L.  [-0.9,0.9] \\ \hline
                  \multirow{2}{*}{$f_{M,3}/\Lambda^4 (\TeV^{-4})$}&68$\%$ C.L. [-5.0,5.0] & 68$\%$ C.L. [-1.2,1.2] \\ 
                      & 95$\%$ C.L. [-7.9,7.9]   &  95$\%$ C.L.  [-1.9,1.9]\\ \hline
                  \multirow{2}{*}{$a_{0}^{W}/\Lambda^2 (\TeV^{-2})$}&68$\%$ C.L. [-1.1,1.1] &68$\%$ C.L. [-0.3,0.3] \\ 
                 &  95$\%$ C.L. [-1.8,1.8]      & 95$\%$ C.L. [-0.5,0.5]    \\ \hline
                  \multirow{2}{*}{$a_{C}^{W}/\Lambda^2 (\TeV^{-2})$}&68$\%$ C.L. [-3.3,3.3] &68$\%$ C.L.  [-0.8,0.8] \\ 			  			
                  &95$\%$ C.L. [-5.2,5.2]        & 95$\%$ C.L. [-1.2,1.2]   \\ \hline

  		\end{tabular}
  	\end{center}
  	\caption{68$\%$ and
95$\%$ C.L.  expected  limits on dimension-8 aQGCs assuming only one of them non-zero while the rest are set
to zero. Expected limits on the dimension-6 aQGCs are also presented
from translation of dimension-8 couplings. The limits include both
acceptance scenarios and the 300 $\text{fb}^{-1}$ IL.} \label{table:quarticlim}
  \end{table}

\begin{figure}
	\begin{center}
		\vspace{1cm}
	\resizebox{0.45\textwidth}{!}{\includegraphics{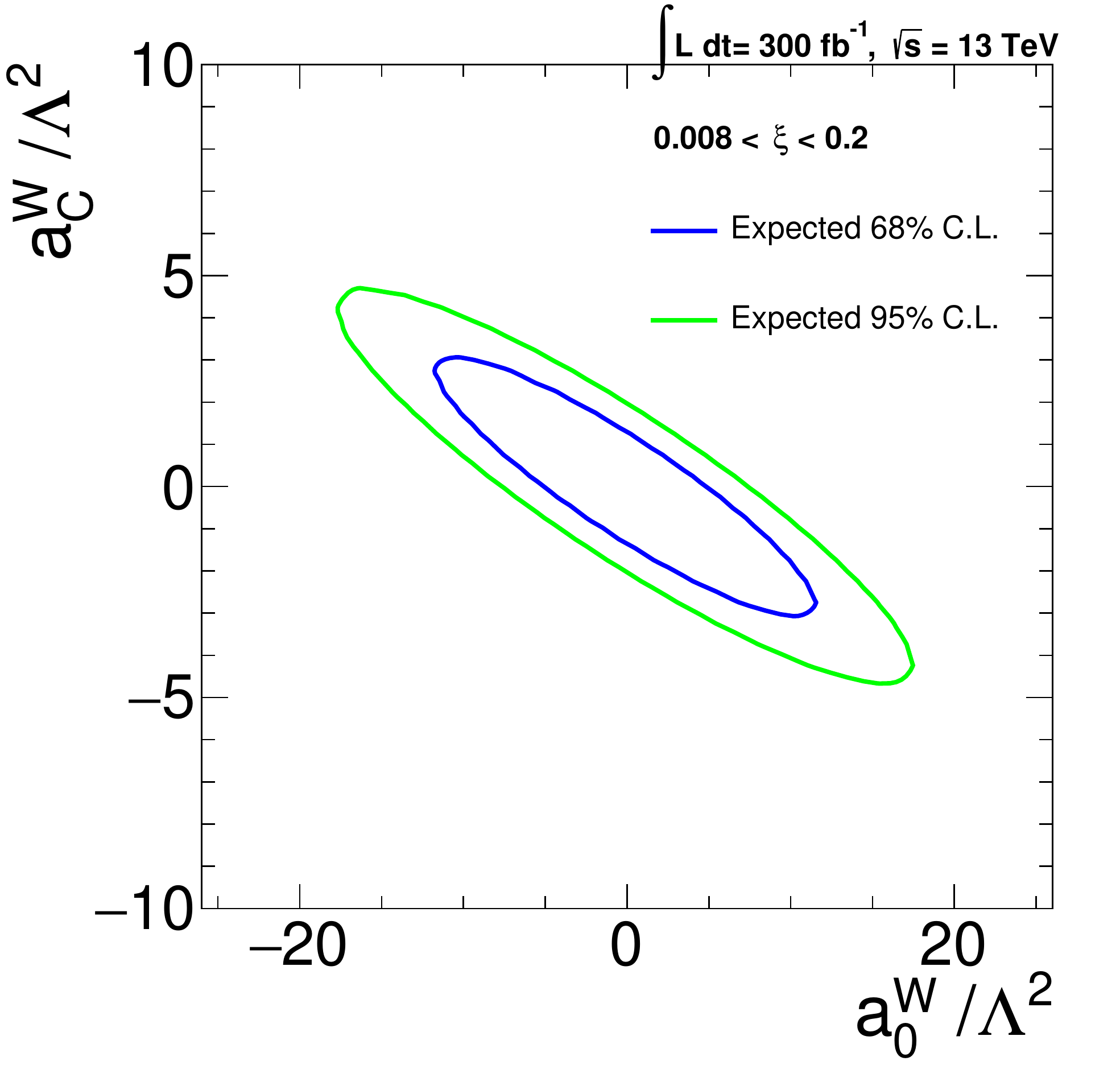}}  
		\resizebox{0.45\textwidth}{!}{\includegraphics{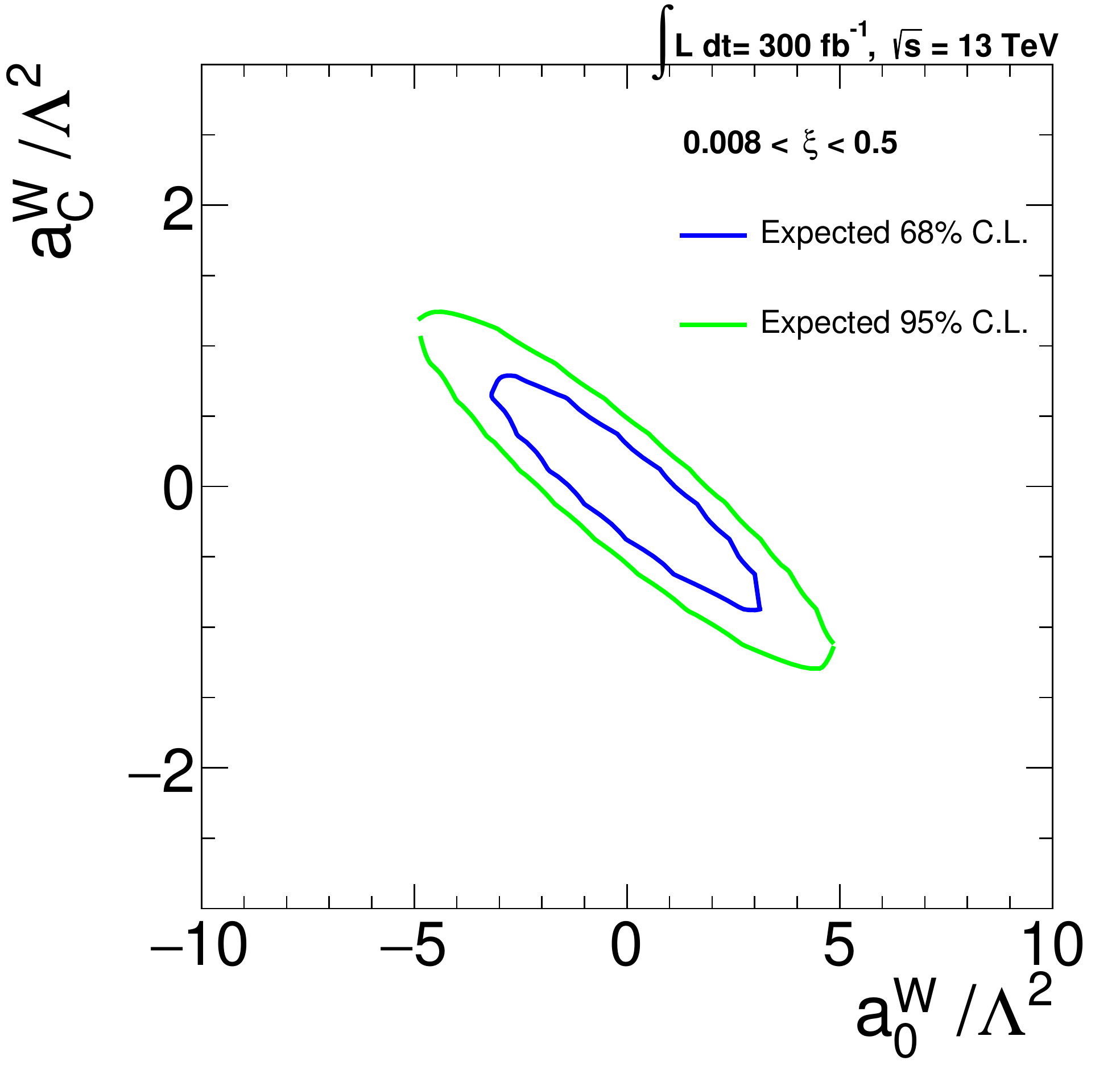}}  

		\caption{ Two-dimensional 68$\%$ and 95$\%$ C.L.
                  expected allowed regions of  $a_{0}^{W}$ and $a_{C}^{W}$ for
acceptance of 0.008$ < \xi < $0.2 (left) and 0.008$ < \xi < $0.5
(right), assuming 300 f$\text{b}^{-1}$ IL.}\label{2dtgbqlimit}
	\end{center}
\end{figure}

As it was explained in  section~\ref{eff:lag} due to similar Lorentz structure of
dimension-8 and dimension-6 operators they can be expressed in terms
of each other which is shown in equation~\ref{8to6trans}. 
	 Using this relation we translated  limits on dimension-8  $f_{M,0,2}$ and $f_{M,1,3}$ anomalous couplings to the expected limit on  dimension-6  $a_{0}^{W}$ and $a_{C}^{W}$ anomalous couplings which are shown in Table~\ref{table:quarticlim} for both assumed acceptance regions of protons. 
In addition, we  calculated the  expected allowed two-dimensional
regions for $a_{0}^{W}$ and $a_{C}^{W}$ by generating the signal sample that
includes the simultaneous variation of $f_{M,0,1,2,3}$ using the re-weighting approach implemented in the \texttt{MadGraph5\_aMC@NLO} package ~\cite{{Alwall:2011uj},{Alwall:2014hca}}.
By scanning the $f_{M,0,1,2,3}$ simultaneously over 400 points and
translation of the expected limit on the $a_{0}^{W}$ and $a_{C}^{W}$ we
obtain the two-dimensional 68$\%$ and 95$\%$ C.L. expected limit
assuming the 300 f$\text{b}^{-1}$ IL shown in
Figure~\ref{2dtgbqlimit} left and right for the acceptance regions of
 0.008$ < \xi < $0.2 and 0.008$ <\xi < $0.5, respectively.

\section{ Summary and remarks}\label{sec:conclu}

 For the first time, the potential of the LHC to measure the CEP of
$W^{+}W^{-}\gamma$ as well as the sensitivity of this process to the 
aTGCs and aQGCs, in the fully leptonic decay channel of W bosons,  is
explored. In contrast to the small predicted cross section of 
$W^{+}W^{-}\gamma$ CEP, this process is highly sensitive to multiple gauge boson
couplings as the tree level diagrams made of purely gauge bosons.

To assess this goal first the feasibility of the LHC to measure the SM
$W^{+}W^{-}\gamma$  production via quasi-real photon-photon scattering
is investigated. The detailed understanding of final state objects both
in the central and FDs are essential to
distinguish signal process from the backgrounds. 
Signal events suffer from two major sources of background that arise from the other CEP processes with the common final state particles and inclusive processes which are
coincided with the pile-up protons. 
Therefore, the presence of forward
detectors with high resolution on momentum and arrival time of protons
is vital to suppress background contributions. To preserve the
optimum amount of signal also having maximum rejection of backgrounds we
introduce three categories of selection cuts. The first set of cuts aims to
keep the least number of objects needed to reconstruct the signal in
the central detector. The second one reflects the acceptance limitations of FDs
to tag intact protons. The final category of cuts benefits from
the high kinematical correlation of central final state objects with the
scattered protons detected by the FDs. The first and third type of
cuts are very efficient for other CEP backgrounds such as
$e^{+}e^{-}\gamma, \mu^{+}\mu^{-}\gamma, \tau^{+}\tau^{-}\gamma$
  processes while all three categories of criterion are useful to reject
  the inclusive backgrounds which occur simultaneously with
  pile-up protons. The contribution of the latter background will grow as
  the mean number of pile-up increase in the high luminosity
  condition of the LHC. We evaluate the probability of tagging 
  protons on FDs considering the energy acceptance and time of flight
  resolution w.r.t the vast range of mean pile-up scenarios
  from 10$-$140. The obtained probabilities can be used for any other studies aiming to estimate
  the contribution of inclusive backgrounds coincides with the pile-up
  protons. Then we estimate the amount of data that is needed
  to have strong evidence of SM predicted $W^{+}W^{-}\gamma$ central
  exclusive process and finally the
  observation of this process.
  \begin{figure}[H]
  	\begin{center}
  		\vspace{1cm}
  		\resizebox{0.47\textwidth}{!}{\includegraphics{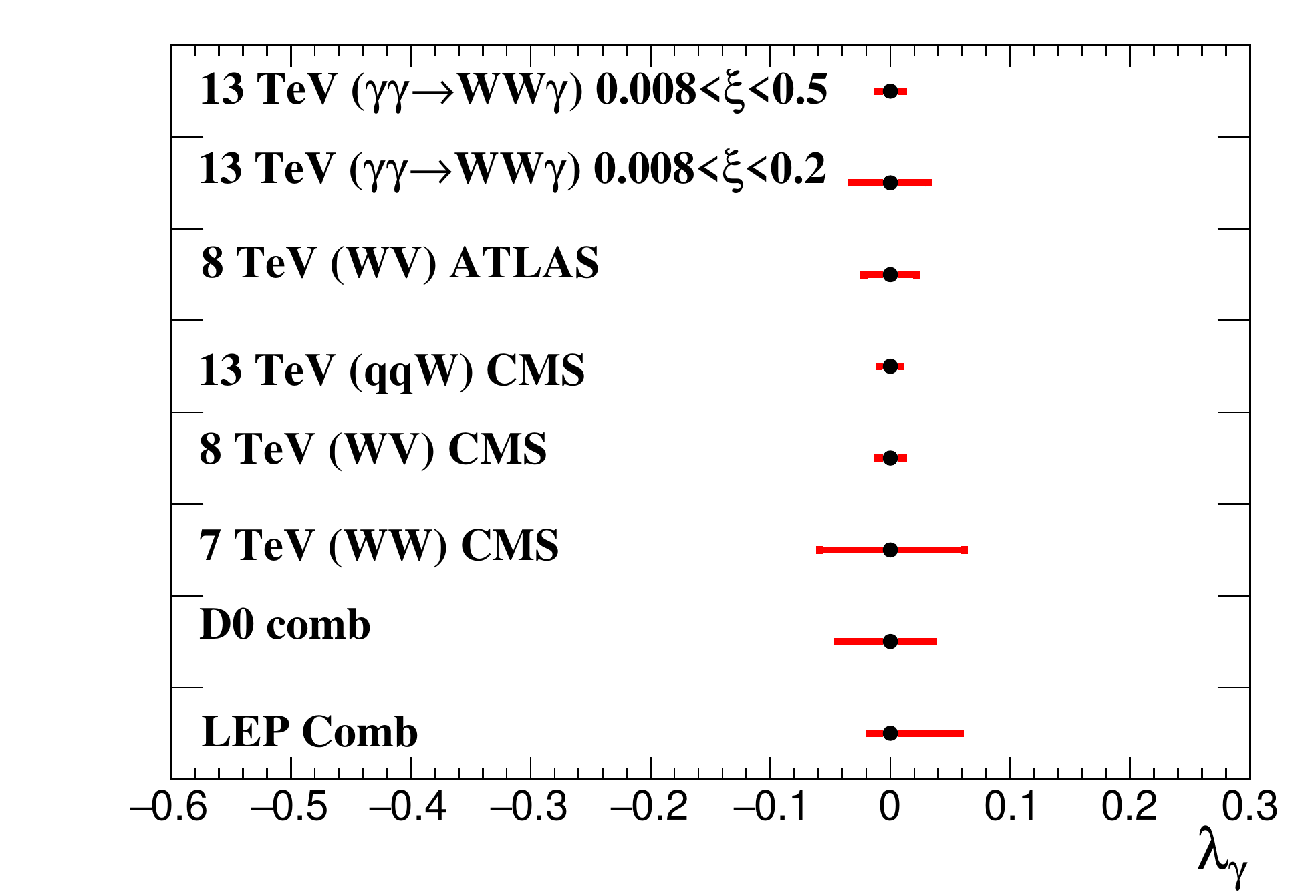}}
  		\resizebox{0.47\textwidth}{!}{\includegraphics{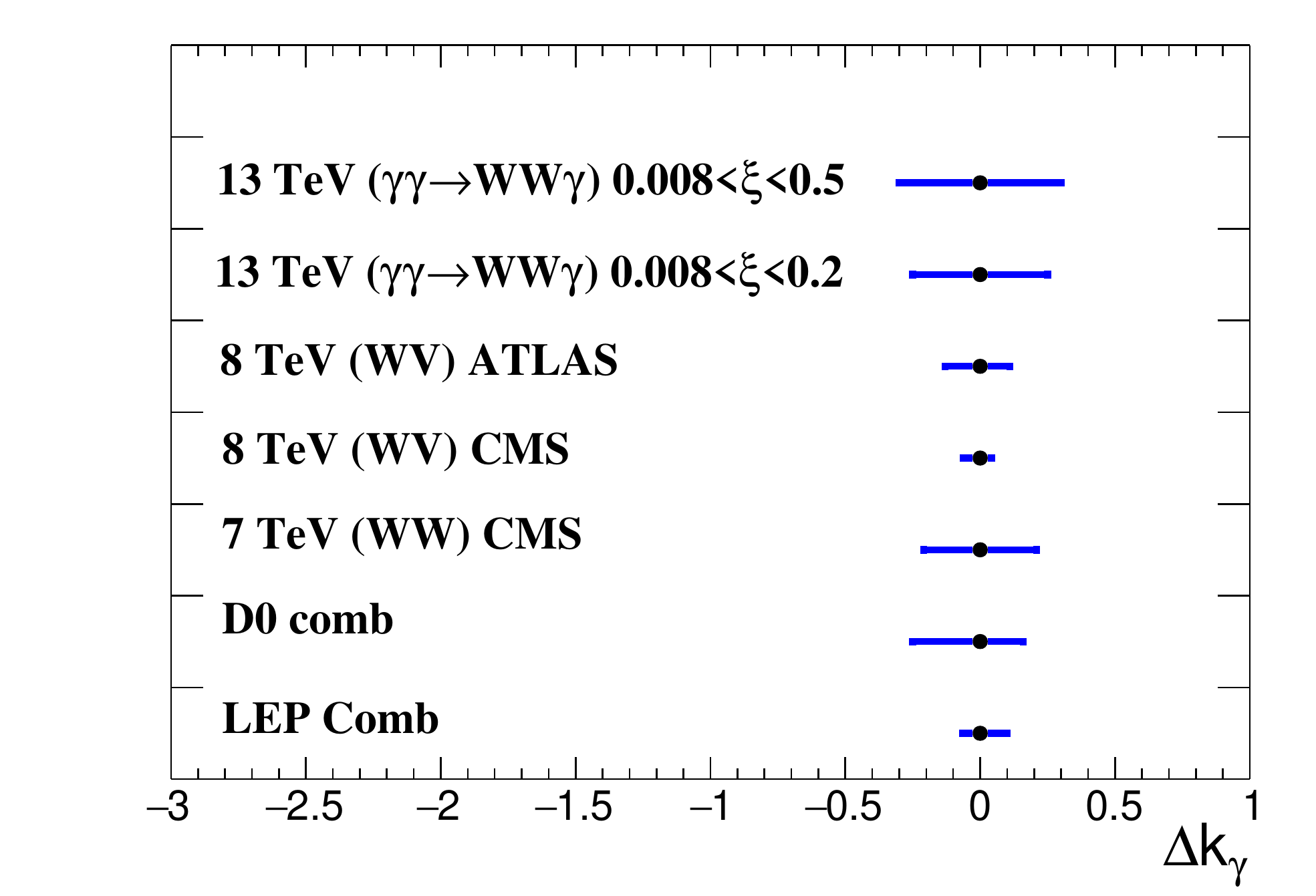}}  
  		\resizebox{0.47\textwidth}{!}{\includegraphics{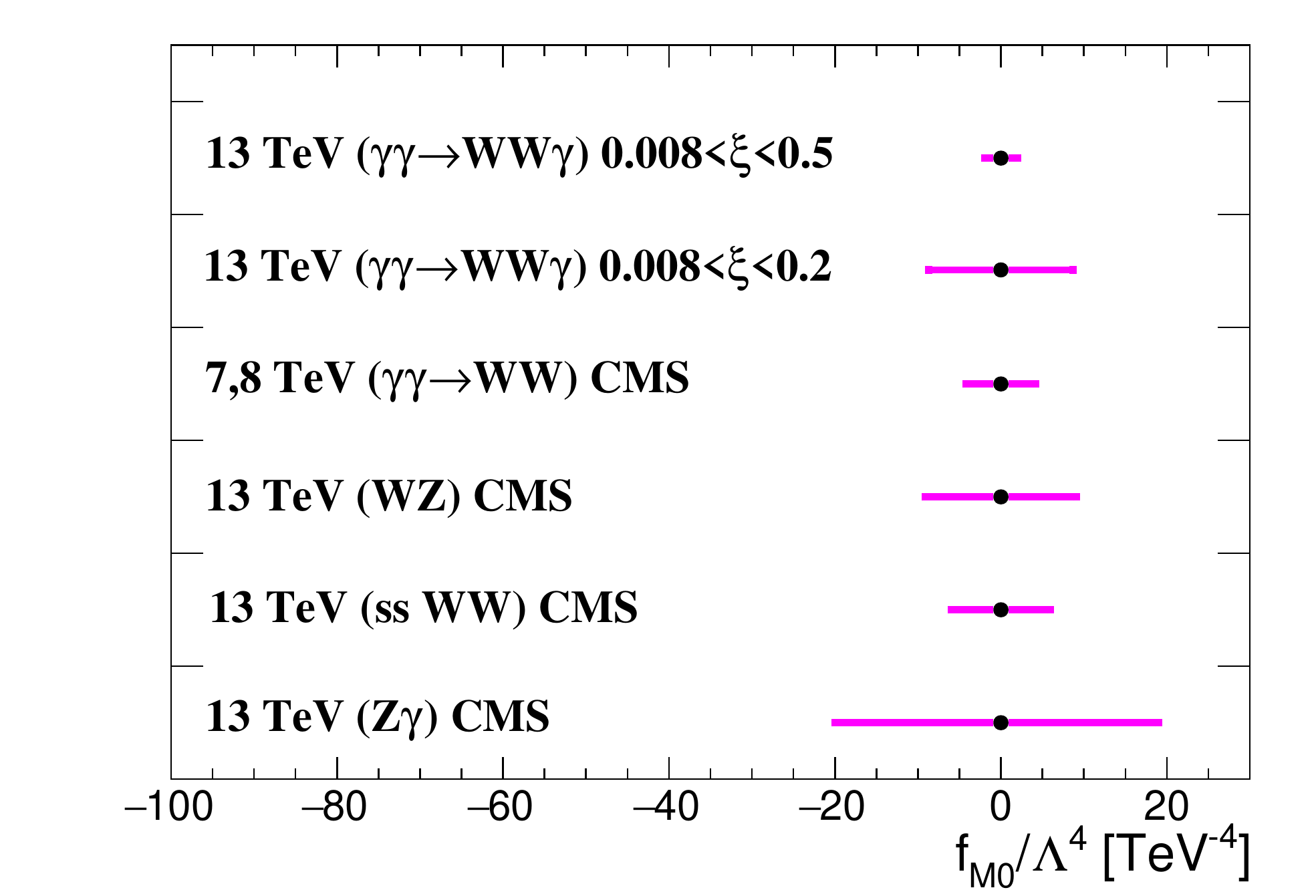}}  
  		\resizebox{0.47\textwidth}{!}{\includegraphics{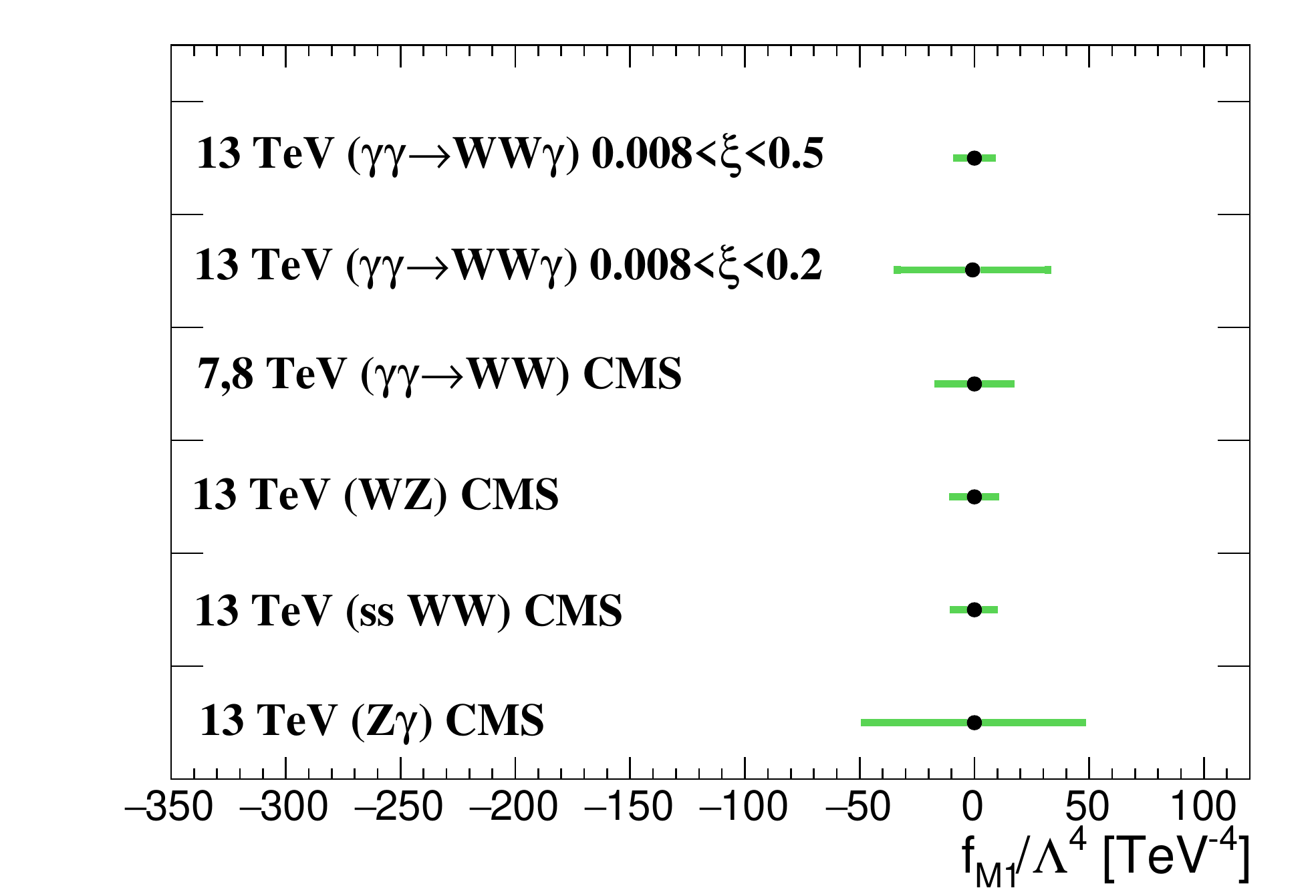}}  
  		\resizebox{0.47\textwidth}{!}{\includegraphics{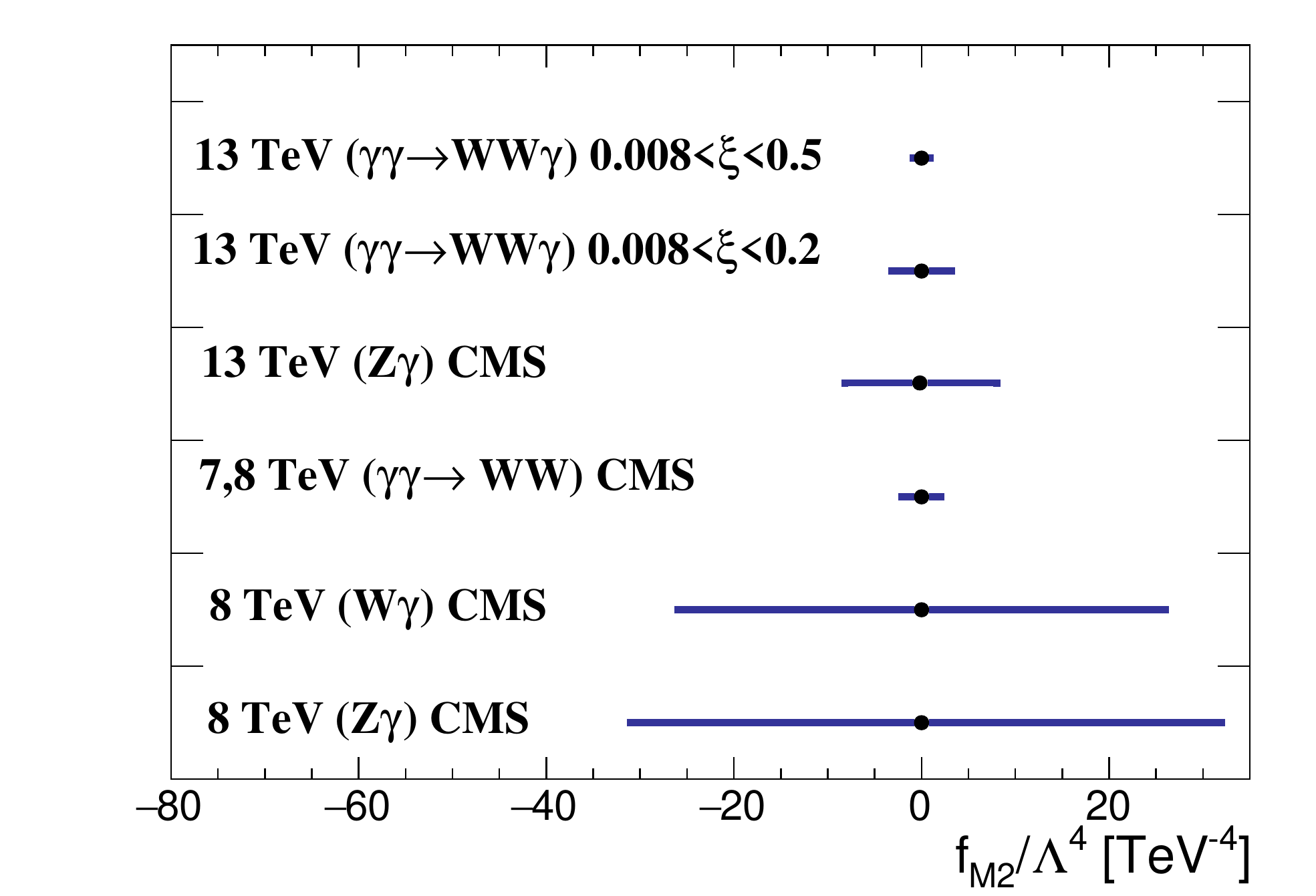}}  
  		\resizebox{0.47\textwidth}{!}{\includegraphics{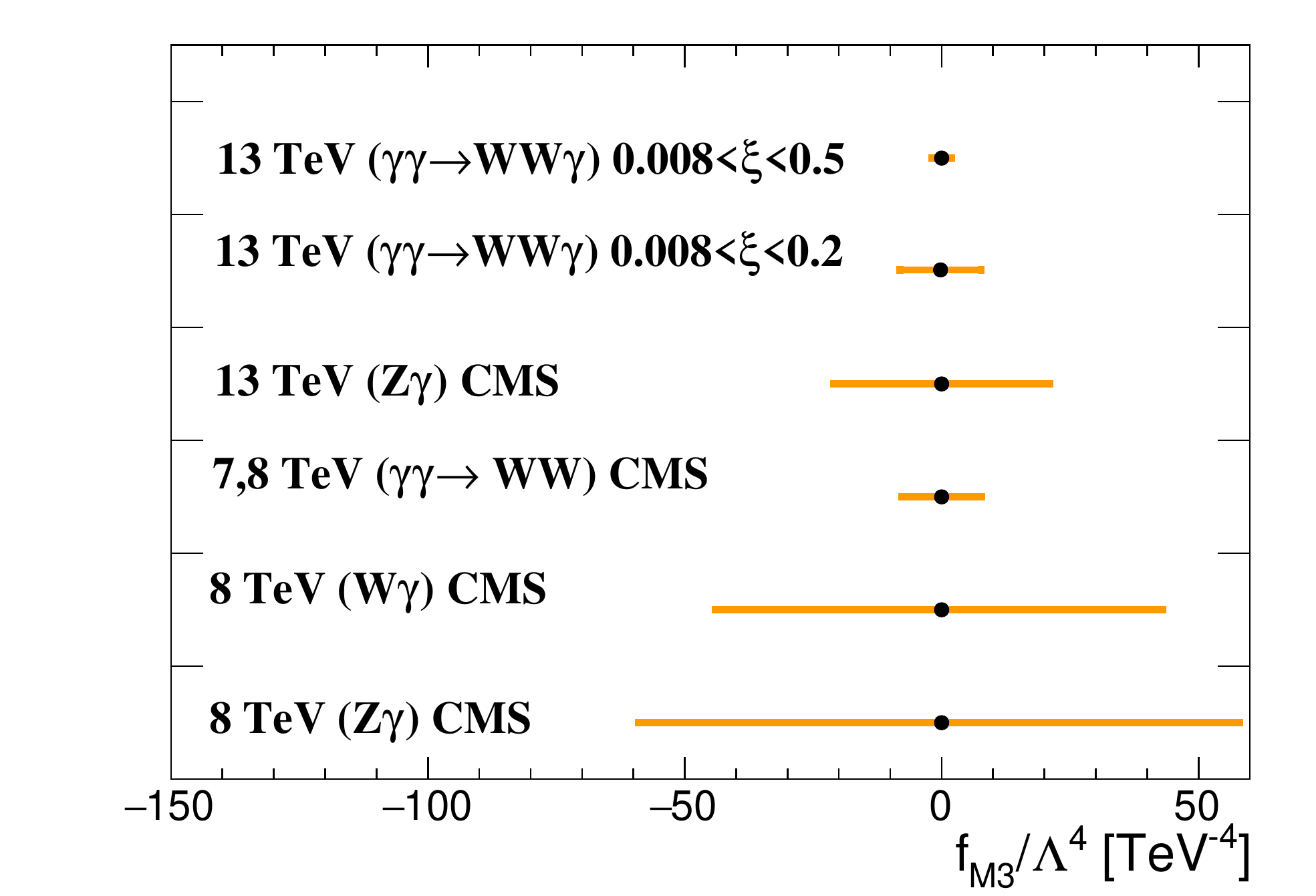}}  
  		
  		
  		\caption{  Comparison of expected 95$\%$ C.L. limit on
  			aTGCs and aQGCs obtained from
  			$\gamma\gamma\rightarrow WW\gamma$ analysis with
  			center-of-mass energy of 13 TeV with current
  			experimental observed limits~\cite{Schael:2013ita:lepcomb,Abazov:2012ze:d0comb,Chatrchyan:2013yaa:cmsww7tev,Sirunyan:2017bey:wvcms8,Aaboud:2017cgf:wv8atlas,Sirunyan:2019dyi:w13cms,Khachatryan:2016mud:exclwwcms,Sirunyan:2019ksz:wzcms13,Sirunyan:2017ret:ssww13cms,Khachatryan:2016vif:wgamma8tevcms,Sirunyan:2020tlu:zgacms13} }\label{summarytqgc}
  	\end{center}
  \end{figure}
  In the second part, we  estimate the power of this process to
  probe the aTGCs and aQGCs. In this regard, the $W^{+}W^{-}\gamma$ CEP
  also counts as irreducible background for the signal with anomalous
  couplings. As these anomalous couplings usually  are emanated from momentum dependent
  terms in the effective Lagrangian. Therefore, the selection cuts introduced in the previous
  part upgraded to obtain the optimum signal region in the
  high momentum phase space. To study the new couplings, we
  consider the Lagrangian based on anomalous coupling approach for
  aTGCs. For aQGCs we employ dimension-8 effective terms that
  contribute to the $WW\gamma\gamma$ vertex. Then the expected limits
  are translated to the two dimension-6 operators contribute to the
  aQGCs. The expected 68$\%$ and 95$\%$ C.L. limit for all anomalous
  couplings are calculated individually. The two-dimensional limits
  are also extracted by obtaining the signal yields when two
  parameters vary 
simultaneously.
   All the limits are expressed in
  two considered acceptances of 0.008$ < \xi < $0.2 and 0.008$ < \xi <
  $0.5 for protons. \par
We have compared the obtained limits of this analysis with the current
experimental bounds on aTGCs and aQGCs  \cite{Schael:2013ita:lepcomb,Abazov:2012ze:d0comb,Chatrchyan:2013yaa:cmsww7tev,Sirunyan:2017bey:wvcms8,Aaboud:2017cgf:wv8atlas,Sirunyan:2019dyi:w13cms,Khachatryan:2016mud:exclwwcms,Sirunyan:2019ksz:wzcms13,Sirunyan:2017ret:ssww13cms,Khachatryan:2016vif:wgamma8tevcms,Sirunyan:2020tlu:zgacms13} which are shown in 
Figure~\ref{summarytqgc}. Left-top plot shows this
process is highly sensitive to the $\lambda_{\gamma}$ while right-top plot indicates the obtained limits on $\Delta \kappa_{\gamma}$ are not competitive to the current bounds. This is partly because the high invariant mass cuts define the signal region while the
$\Delta \kappa_{\gamma}$ coupling only alters the normalization of the SM
process. Regarding the aQGCs, this analysis shows very good sensitivity
to these couplings as it is obvious from  left-middle, right-middle, left-bottom,
and right-bottom plots which compare the expected limits on the
$f_{M,0,1,2,3}$  to the current experimental
observed limits by the CMS and ATLAS experiments,
respectively. These plots show using
  $\gamma\gamma\rightarrow W^{+}W^{-}\gamma$ process one could obtain considerable improvement, especially on $f_{M,2}$ and $f_{M,3}$ couplings. Also, sensitivity on all four couplings is competitive with the
inclusive $\gamma\gamma\rightarrow W^{+}W^{-}$ process measured by the
CMS experiment~\cite{Khachatryan:2016mud:exclwwcms}. In summary, this study shows the $W^{+}W^{-}\gamma$ CEP is
very effective to probe the aQGCs and could be used by the current LHC
experiments as a sensitive as well as a complementary channel to probe the multi-gauge boson couplings predicted in the SM.


%
{\bf Acknowledgments:}

We are thankful to Christophe Royon and Mojtaba Mohammadi Najafabadi for insightful discussions and their helpful comments on the manuscript.

\end{document}